\documentclass{aa}
\usepackage{txfonts}
\usepackage{natbib}
\usepackage{epsfig,latexsym,graphicx,epsf,subfigure}
\usepackage{threeparttable,hhline,longtable}
\usepackage{deluxetable} 

\newcommand{\chisq}{\chi^2}
\newcommand{\eqref}[1]{(\ref{#1})}

\def\lsim{\raise0.3ex\hbox{$<$}\kern-0.75em{\lower0.65ex\hbox{$\sim$}}}
\def\gsim{\raise0.3ex\hbox{$>$}\kern-0.75em{\lower0.65ex\hbox{$\sim$}}}
                                                               
\begin{document}

\title{\bf Restframe $I$-band Hubble diagram for type~Ia supernovae up
to redshift $z \sim$~0.5\thanks{Based in part on observations taken at
the European Southern Observatory using the ESO Very Large Telescope
on Cerro Paranal (ESO Program 265.A-5721(B)). Based in part on
observations made with the NASA/ESA Hubble Space Telescope, obtained
at the Space Telescope Science Institute, which is operated by the
Association of Universities for Research in Astronomy, Inc., under
NASA contract NAS 5-26555. These observations are associated with
program GO-8346. Based in part on data collected from the
Canada-France-Hawaii Telescope Corporation, which is operated by the
National Research Council of Canada, le Centre National de la
Recherche Scientifique de France, and the University of Hawaii.  Based
in part on observations taken at the W.M. Keck Observatory, which is
operated as a scientific partnership among the California Institute of
Technology, the University of California and the National Aeronautics
and Space Administration. The Observatory was made possible by the
generous financial support of the W.M. Keck Foundation.}}

\author{S.~Nobili\inst{1,2},
R.~Amanullah\inst{2},
G.~Garavini\inst{1,2},
A.~Goobar\inst{2},
C.~Lidman\inst{3},
V.~Stanishev\inst{2},
G.~Aldering\inst{4},
P.~Antilogus\inst{1},
P.~Astier\inst{1},
M.~S.~Burns\inst{5},
A.~Conley\inst{4,6},
S.~E.~Deustua\inst{7},
R.~Ellis\inst{8},
S.~Fabbro\inst{9},
V.~Fadeyev\inst{4},
G.~Folatelli\inst{2},
R.~Gibbons\inst{4},
G.~Goldhaber\inst{4,6},
D.~E.~Groom\inst{4},
I.~Hook\inst{10},
D.~A.~Howell\inst{4},
A.~G.~Kim\inst{4},
R.~A.~Knop\inst{11},
P.~E.~Nugent\inst{4},
R.~Pain\inst{1},
S.~Perlmutter\inst{4,6},
R.~Quimby\inst{4},
J.~Raux\inst{1},
N.~Regnault\inst{1,4},
P.~Ruiz-Lapuente\inst{12},
G.~Sainton\inst{1},
K.~Schahmaneche\inst{1},
E.~Smith\inst{11},
A.~L.~Spadafora\inst{4},
R.~C.~Thomas\inst{4}, and
L.~Wang\inst{4}\\
(THE SUPERNOVA COSMOLOGY PROJECT)\\[12pt]
}

\institute{
LPNHE, CNRS-IN2P3, University of Paris VI \& VII, Paris, France  \and
Department of Physics, Stockholm University,  Albanova 
       University Center, S-106 91 Stockholm, Sweden \and
European Southern Observatory, Alonso de Cordova 3107, 
       Vitacura, Casilla 19001, Santiago 19, Chile  \and
E. O. Lawrence Berkeley National Laboratory, 1 Cyclotron Rd., Berkeley, CA 94720, USA   \and
Colorado College, 14 East Cache La Poudre St., Colorado Springs, CO 80903  \and
Department of Physics, University of California 
Berkeley, Berkeley, 94720-7300 CA, USA   \and
American Astronomical Society,  2000 Florida Ave, NW, 
Suite 400, Washington, DC, 20009 USA \and
California Institute of Technology, E. California 
Blvd, Pasadena,  CA 91125, USA \and
CENTRA-Centro M. de Astrof\'{\i}sica and Department of 
Physics, IST, Lisbon, Portugal \and
Department of Physics, University of Oxford, Nuclear 
\& Astrophysics Laboratory,  Keble Road, Oxford, OX1 3RH, UK \and
Department of Physics and Astronomy, Vanderbilt 
University, Nashville, TN 37240, USA \and
Department of Astronomy, University of Barcelona, 
Barcelona, Spain}

\authorrunning{Nobili et al.}
                                                                               
\offprints{S.~Nobili , serena@lpnhep.in2p3.fr}
                                               
\date{Received ...; accepted ...}

\abstract{ We present a novel technique for fitting restframe $I$-band
  light curves on a data set of 42 Type~Ia supernovae (SNe~Ia). Using
  the result of the fit, we construct a Hubble diagram with 26 SNe
  from the subset at $0.01< z<0.1$. Adding two SNe at $z\sim0.5$
  yields results consistent with a flat $\Lambda$-dominated
  ``concordance universe''
  ($\Omega_M,\Omega_\Lambda$)=(0.25,0.75). For one of these,
  SN~2000fr, new near infrared data are presented.  The high redshift
  supernova NIR data are also used to test for systematic effects in
  the use of SNe~Ia as distance estimators. A flat, $\Lambda=0$,
  universe where the faintness of supernovae at $z\sim0.5$ is due to
  grey dust homogeneously distributed in the intergalactic medium is
  disfavoured based on the high-z Hubble diagram using this small
  data-set.  However, the uncertainties are large and no firm
  conclusion may be drawn.  We explore the possibility of setting
  limits on intergalactic dust based on $B-I$ and $B-V$ colour
  measurements, and conclude that about 20 well measured SNe are
  needed to give statistically significant results.  We also show that
  the high redshift restframe $I$-band data points are better fit by
  light curve templates that show a prominent second peak, suggesting
  that they are not intrinsically underluminous.
  } \maketitle

\section{Introduction}

Observations of Type~Ia supernovae in the restframe $B$-band at
redshifts of $z \sim 0.5$ and above have shown that they are best fit
by a cosmological model that includes a cosmological constant or some
other form of dark energy
\citep{nature,garnavich,Riess,schmidt,perlmutter42,tonry,rob03,riess04}. The
evidence for dark energy is supported by cross-cutting cosmological
results, such as the measurement of the cosmic microwave background
anisotropy, which indicates a flat universe
\citep{DeB,Jaffe,Sie03,wmap}; the evolution in the number density of
X-ray emitting galaxy clusters \citep{Borgani,Henry,Schuecker} and
galaxy redshift surveys \citep{Efstathiou}, which indicate that
$\Omega_{\rm M} \approx 0.3$. Taken together, these independent
measurements suggest a concordance universe with ($\Omega_{\rm
M}$,$\Omega_{\Lambda}$)$\cong$(0.25,0.75). However, the SN~Ia Hubble
diagram remains the most direct approach currently in use for studying
cosmic acceleration, and, thus, possible systematic effects affecting
the observed brightness of Type~Ia supernovae should be carefully
considered, such as
uncorrected host galaxy extinction (see e.g. \citet{RowanRobinson}),
dimming by photon-axion mixing over cosmological distances
\citep{CKT,deffayet,edvard,linda} and extinction by intergalactic grey dust
\citep{aguirre1999a,aguirre1999b,2003JCAP...09..009M}.
Some of these have already been addressed in previous SCP
publications, see e.g. \citet{perlmutter97,nature,sullivan,rob03}.

Determining cosmological distances through Type~Ia supernova fluxes
at longer restframe wavelengths offers potential advantages, e.g. less
extinction by dust along the line of sight, either in the host galaxy
or in the intergalactic medium. On the other hand, the ``standard
candle'' properties at these wavelengths and the possibility of additional
systematic effects need to be investigated.

In the restframe $I$-band, the uncertainties in extinction
corrections are significantly smaller than those in the $B$-band. For
example, for Milky-Way type dust ($R_V\sim 3$) the ratio of extinction
for the two bands is sizable, $A_B/A_I \sim 2 - 3$. In general, the
extinction corrections become less dependent on our knowledge of 
intrinsic supernova colours and dust properties.

SNe~Ia $I$-band light curves typically show a second peak 15-30 days
after the first maximum. It has been suggested that the intensity and
time-difference between the first and second $I$-band peaks are
related to the intrinsic luminosity of the Type~Ia SNe, appearing
later and more evident for normal Type~Ia and earlier and fainter for
underluminous ones \citep{Hamuy1996,wang}. Thus, building $I$-band
light curves for Type~Ia supernovae offers the possibility of probing
brightness evolution.

The scope of this work is to test the feasibility of using the
restframe $I$-band for cosmological distance measurements, using data
available to date, and to assess the importance of observing in this
wavelength range for future samples of SNe. For that purpose, we
develop a template fitting technique, which we apply to 42 nearby
SNe~Ia, to estimate the first ($I_{\rm max}$) and second ($I_{\rm
sec}$) $I$-band light curve peaks.  We use the fitted $I_{\rm max}$ of
26 of these SNe~Ia, which are in the Hubble flow, together with two
high redshift SNe~Ia: SN~2000fr, at redshift $z=0.543$, for which new
infrared data are presented, and SN~1999ff, at $z=0.455$, available in
literature \citep{tonry}, to build a Hubble diagram reaching out to
$z\sim0.5$.

The properties of the second peak in the restframe $I$-band light
curves are investigated. Furthermore, additional colour information is
used to test for extinction by non-conventional dust for the
$z\sim0.5$ supernovae.  In a related work, \citet{riess99q}
used $B-I$ colours of SN~1999Q, in the same redshift range. This SN,
however, is not included in our analysis, since we find
inconsistencies with the published data (see Section~\ref{sec:99q}).

\section{I-band light curve fitting}
\label{sec:ibandfit}

The second light curve peak seen in $I$-band for nearby Type~Ia SNe
varies in strength and position with respect to the primary maximum.
This complicates the use of a singly parametrised $I$-band template,
such as those currently applied in the $B$- and $V$-band, (see
e.g. \citet{perlmutter97,gerson} for an example of the timescale
stretch factor approach), for light curve fitting.

\citet{contardo} proposed a model composed of as many as 4 functions
for a total of 10 parameters in order to fit all $UBVRI$-bands.  Their
method used two Gaussian functions to fit the two peaks, together with
a straight line to fit the late time decline and an exponential factor
for the pre-max rising part of the light curve.  In this way, it is
possible to describe Type~Ia SNe light curves over a wide range of
epochs and in all optical bands, though, as the authors recognise, it
does not represent accurately the second peak in the $I$-band due to
the influence of the linear decline. However, the main disadvantage of
their method, for our purpose, is the large number of free parameters,
which requires very well sampled light curves.

We have therefore developed a method for fitting $I$-band light curves
using five free parameters and one template\footnote{The $B$-band
template in \citet{Bnugent} has been used because we found that it
describes well the data when used with the method developed here. We
note that there are no physical reasons for choosing the $B$-band over
other bands or other kinds of templates.} used twice to describe the
two peaks.  As our goal is only to measure the position and amplitude
of the two peaks, we limit the fit to 40 days after maximum,
neglecting the late time decline. Our fitting procedure can be
summarised as follows: one template is used to fit the time ($t_1$)
and the first peak magnitude ($I_1$), together with a stretch factor
($s_I$), which is also applied to the second template shifted in time
to fit the time ($t_2$) and magnitude of the second peak ($I_2$). The
underlying function is
$$I = I_1{\cal T}(s_I(t-t_1)) + I_2 {\cal T}(s_I(t-t_2))$$ where
${\cal T}$ is the template.  The five parameters fitted are thus:
$\{t_1,t_2,I_1,I_2,s_I\}$, (see Table \ref{legend}). A similar
approach is also proposed by \citet{wang} who call it
``super-stretch'' to emphasise its extension of the stretch approach.

The use of this function in place of the one described by
\citet{contardo}, reduces the number of free parameters by a factor of
two.  Implicitly, we have thus assumed that the rising part of the
$I$-band is the same as in the template used, i.e. the $B$-band.  As
we will see, this assumption is not always true.  Note that, unless
otherwise specified, the supernova phase always refers to the time
relative to restframe $B$-band light curve maximum.

\begin{table}[htb]
\begin{center}
\begin{tabular}{ll}\hline\hline
$t_1$ & time of the peak of the first ${\cal T}$ template \\
$I_1$ & peak magnitude of the first ${\cal T}$ template \\
$t_2$ & time of the peak of the second ${\cal T}$ template\\
$I_2$ & peak magnitude of  the second ${\cal T}$ template\\
$s_I$ & stretch factor of the time axis\\
\hline
$t_{\rm max}$ & time of the first $I$ light curve peak\\
$I_{\rm max}$ & first $I$ light curve peak magnitude\\
$t_{\rm sec}$ & time of the second $I$ light curve peak\\
$I_{\rm sec}$ & second $I$ light curve peak magnitude\\
\hline
\end{tabular}
\caption{Summary of the parameters used in this work to describe the
$I$-band light curve. The first five parameters are determined by
fitting the data (see text for details). The next four parameters are
determined from the first set and are the actual time and peak values
of the light curve.}
\label{legend}
\end{center}
\end{table}

\subsection{The low-redshift data set}
\label{sec:data}
We applied this method to fit a sample of local SNe~Ia for which both
$B$ and $I$-band data are available in the literature. The SNe
primarily come from the Calan/Tololo \citep{Hamuy1996}, CfA
\citep{Riess22} and CfA2 \citep{jha} data sets. Data from three other
well studied individual supernovae were also included: SN~1989B
\citep{89b}, SN~1994D \citep{94d} and the underluminous supernova
SN~1991bg. We have used two data sets in restframe $I$-band for
SN~1991bg, one published by \citet{Filippenko92} with quite good
coverage from about 3 days after $B$-band maximum light to +60 days,
and another published by \citet{leibundgut93} with four data points,
the first of which is at the time of $B$-band maximum. The agreement
between the two data sets was assessed by comparing the measurements
taken at the same date, i.e. JD=2448607, where we found a difference
of 0.06 mag.  We take this as an estimate of the measurement
uncertainty in data of \citet{leibundgut93} as no uncertainties are
reported in that work.

\subsection{Fitting method and results}
\label{sec:fitting}
Only supernovae with at least 6 $I$-band data points and time coverage
constraining both peaks were selected for light curve fitting. This
resulted in a total of 42 SNe. Table~\ref{Ifit_results} lists the
parameters resulting from the fitting procedure.  Since the dominant
uncertainties are symmetric in units of flux, we performed the fit in
flux rather than magnitudes. The parameters given in
Table~\ref{Ifit_results} are transformed into magnitudes.

Prior to fitting, all data points were $k$-corrected to restframe
$I$-band as in \citet{KimGoobarPerlmutter96} and \citet{Bnugent},
assuming a Bessell $I$-band filter transmission curve
\citep{Bessellfilters} and time information from the available
$B$-band data. A new spectroscopic template, which is a slightly
modified version of the template found in \citet{Bnugent}, was built
for computing the $k$-corrections. We have preserved the SED from the
UV through the Si II 6150\AA\ feature, following \citet{nobili2003},
but red ward of that we have incorporated additional spectra from the
Supernova Cosmology Project (SCP) Spring 1999 search
\citep{Ald,nugent2000} to improve this region as the original template
was sparse and required a lot of interpolation. A potential source of
systematic uncertainty in the $k$-corrections is due to the wide Ca IR
triplet absorption feature, found to vary considerably among Type ~Ia
supernovae \citep{Strolger2002,Bnugent}. We have estimated this
systematic uncertainty as a function of redshift for $0.01 <z <0.1$ by
computing the $k$-correction for diverse nearby SNe~Ia at different
epochs. The dispersion in the $k$-correction increases with redshift,
reaching 0.05 mag. at $z=0.1$.  We take this as a conservative
estimate of the uncertainty in all $k$-corrections in this wavelength
region.

Note that the values of $I_2$ reported in Table~\ref{Ifit_results} are
not the actual magnitudes of the secondary peak , $I_{\rm sec}$, but a
parameter indicating the size of the contribution of the second
template to the overall I-band light curve.

In Fig.~\ref{plot9}, all the fitted light curves are shown.  They are
sorted in chronological order, except for the two very underluminous
supernovae: SN~1991bg and SN~1997cn, displayed at the bottom of the
figure. As the date of the $B$-band maximum for SN~1997cn is unknown, the
origin of the time axis was set to the epoch ($JD=2450597.75$) when
this supernova was first observed. Note that the second peak of
underluminous supernovae is almost completely absent, resulting in a
value of $I_2$ $\sim$ 2.5 to 3 magnitudes fainter than $I_1$.

Our sample includes SNe that are classified as spectroscopically
peculiar, showing similarities with the over-luminous SN~1991T
\citep{filippenko91T,ruiz91T,phillips91T}. These are SN~1995bd,
SN~1997br, SN~1998ab, SN~1998es, SN~1999aa, SN~1999ac, SN~1999dq and
SN~1999gp  \citep{li,Howell,gabri99aa}. One supernova, SN~1993H,
was reported to show similarities with the spectrum of the peculiar
underluminous SN~1986G \citep{1993IAUC.5723R...1H}.  However, as we
will see in this work, we do not find all of these to show
peculiarities in their $I$-band light curve shape when compared to
spectroscopically normal SNe~Ia. Recently,
\citet{2003astro.ph.12626K}, built the Hubble diagram for SNe~Ia in
infrared $J,H$ and $K$-bands out to $z=0.04$, and reported that three
spectroscopically peculiar SNe, SN~1999aa, SN~1999ac and SN~1999aw, do
not show a behaviour different than that of normal SNe.  With the aim
to assess a greater homogeneity of SNe as standard candles in the
$I$-band than in $B$-band, we choose not to exclude peculiar SNe from
our sample, and instead monitor possible deviant behaviour of these
objects.

Analysing the results of our fits, we found that Type~Ia SNe show a
variety of properties for the $I$-band light curve shape. In
particular we noticed that the light curve could peak between $-3$
days and $+3$ days w.r.t. $B_{\rm max}$, as shown in
Fig.~\ref{histo_params} (left-hand panel). The time of the second
peak, $t_{\rm sec}$ (relative to $B_{\rm max}$), is shown in the
right-hand panel. The distribution of $t_{\rm max}$ is centred at day
$-0.3$ and has a dispersion of $\sigma=1.3$ days. $t_{\rm sec}$ is
centred at 23.6 with a dispersion of $\sigma=4.4$ days.  The result
shown in Fig.~\ref{histo_params} can be compared with the result of
\citet{contardo} (their Fig.~4). Selecting the subsample used by them
we obtain a similar distribution, quite flat and spread over a broad
range, centred around 2 days before $B$-maximum.  However, when more
SNe are added, we obtain the distribution shown in
Fig.~\ref{histo_params}.

The fits have reduced $\chisq$ values (see Table~\ref{Ifit_results})
that are generally around unity, except for a few cases. 

Approximately half of the fits have reduced $\chisq$ values (see
Table~\ref{Ifit_results}) that are around unity. The other half are
either too good or too poor, which either suggests that the published
uncertainties are unreliable or that the template is not a good model
In particular, we note that SN~1994D has a $\chisq/dof \sim
26$. Although it has been suggested that the uncertainties for this
supernova may be underestimated (see \citet{rob03}), the trend in the
residuals shows that this SN is not well described by the model,
indicating the limitations of the fitting function.  As in other cases
we find a systematic trend, especially in the rising part of the light
curves for 6 objects, less than half the supernovae that have
pre-maximum data.  We note however, that a different choice of the
template, selected for fitting the pre maximum data for these 6 SNe,
would fail to fit the rest of the sample, which is well fitted by the
current template. We investigated possible systematic effects in the
fitted light curve maximum due to this, but found no evidence of a
trend in the residuals in the Hubble diagram (see Section\ref{sec:hd})
for these 6 SNe.

While the $\chisq$ gives a measurement of the goodness of the fit, in
the next section we test the robustness and accuracy of the parameter 
estimation in our fitting method, reported in Table~\ref{Ifit_results}.

\begin{table*}[htb]
\begin{center}
\begin{tabular}{lcccccccc}\hline
SN & z & $s_I$ & $t_1$ & $I_1$ & $t_2$ & $I_2$ & $\chisq$ & $N$ \\ 
\hline\hline 
 1989B$^1$  & 0.002  &  1.100 $\pm$  0.126  & -0.590 $\pm$  1.549  & 11.752 $\pm$  0.059  & 23.123 $\pm$  2.079  & 12.513 $\pm$  0.175  &   5.12  & 15 \\
1991bg$^4$  & 0.005  &  1.104 $\pm$  0.034  &  3.274 $\pm$  0.380  & 13.521 $\pm$  0.006  & 28.800 $\pm$  1.123  & 16.506 $\pm$  0.122  &  21.78  & 20 \\
1992al$^4$  & 0.015  &  0.952 $\pm$  0.054  &  1.176 $\pm$  1.016  & 15.039 $\pm$  0.043  & 27.887 $\pm$  1.078  & 15.527 $\pm$  0.069  &   0.59  & 10 \\
1992bc$^4$  & 0.020  &  1.121 $\pm$  0.030  & -1.579 $\pm$  0.146  & 15.639 $\pm$  0.014  & 27.677 $\pm$  0.492  & 16.510 $\pm$  0.026  &  18.78  & 20 \\
1992bg$^4$  & 0.035  &  0.963 $\pm$  0.065  &  0.257 $\pm$  1.929  & 17.543 $\pm$  0.086  & 26.392 $\pm$  2.543  & 17.963 $\pm$  0.060  &   3.18  &  7 \\
1992bh$^4$  & 0.045  &  1.086 $\pm$  0.145  & -0.008 $\pm$  1.029  & 17.899 $\pm$  0.028  & 26.766 $\pm$  2.494  & 18.543 $\pm$  0.116  &   4.16  & 10 \\
1992bo$^4$  & 0.019  &  0.952 $\pm$  0.019  & -0.609 $\pm$  0.162  & 16.064 $\pm$  0.016  & 23.403 $\pm$  0.374  & 16.944 $\pm$  0.042  &  43.91  & 18 \\
1992bp$^4$  & 0.079  &  0.891 $\pm$  0.053  & -0.292 $\pm$  0.710  & 18.962 $\pm$  0.029  & 27.750 $\pm$  1.287  & 19.346 $\pm$  0.084  &  10.14  & 14 \\
1993ag$^4$  & 0.049  &  0.924 $\pm$  0.058  &  1.092 $\pm$  0.863  & 18.384 $\pm$  0.039  & 26.904 $\pm$  2.010  & 18.759 $\pm$  0.060  &   2.57  & 12 \\
 1993H$^4$  & 0.024  &  0.953 $\pm$  0.033  & -2.072 $\pm$  0.960  & 16.664 $\pm$  0.040  & 20.556 $\pm$  0.943  & 17.532 $\pm$  0.048  &  23.61  & 13 \\
 1993O$^4$  & 0.051  &  1.089 $\pm$  0.080  &  0.462 $\pm$  0.917  & 18.197 $\pm$  0.023  & 26.248 $\pm$  1.108  & 18.858 $\pm$  0.034  &  15.97  & 16 \\
1994ae$^3$  & 0.004  &  1.051 $\pm$  0.017  & -1.110 $\pm$  0.142  & 13.383 $\pm$  0.018  & 26.664 $\pm$  0.312  & 14.032 $\pm$  0.042  &  22.65  & 20 \\
 1994D$^2$  & 0.002  &  0.891 $\pm$  0.004  & -1.007 $\pm$  0.043  & 12.177 $\pm$  0.004  & 25.219 $\pm$  0.091  & 12.836 $\pm$  0.008  & 546.17  & 26 \\
 1994M$^3$  & 0.023  &  0.945 $\pm$  0.041  &  0.035 $\pm$  1.262  & 16.513 $\pm$  0.050  & 24.824 $\pm$  1.027  & 17.139 $\pm$  0.063  &  26.73  & 13 \\
 1994T$^3$  & 0.035  &  0.740 $\pm$  0.026  &  2.371 $\pm$  1.319  & 17.458 $\pm$  0.049  & 30.187 $\pm$  0.988  & 17.840 $\pm$  0.053  &   9.22  &  8 \\
1995al$^3$  & 0.005  &  1.158 $\pm$  0.046  & -1.002 $\pm$  0.472  & 13.526 $\pm$  0.022  & 24.696 $\pm$  0.751  & 14.149 $\pm$  0.057  &   8.72  & 16 \\
1995bd$^3$  & 0.016  &  1.166 $\pm$  0.025  & -0.095 $\pm$  0.112  & 16.082 $\pm$  0.012  & 26.516 $\pm$  0.367  & 16.632 $\pm$  0.060  &  20.70  & 16 \\
 1995D$^3$  & 0.007  &  1.267 $\pm$  0.054  & -1.408 $\pm$  0.682  & 13.708 $\pm$  0.026  & 24.763 $\pm$  0.744  & 14.454 $\pm$  0.037  &  10.34  & 25 \\
 1995E$^3$  & 0.012  &  1.026 $\pm$  0.040  &  0.067 $\pm$  0.635  & 15.393 $\pm$  0.024  & 26.340 $\pm$  0.950  & 16.093 $\pm$  0.050  &   6.78  & 14 \\
1996ai$^3$  & 0.003  &  1.115 $\pm$  0.024  & -2.042 $\pm$  0.495  & 13.986 $\pm$  0.022  & 24.947 $\pm$  0.504  & 14.675 $\pm$  0.023  &  48.38  & 10 \\
1996bl$^3$  & 0.036  &  0.942 $\pm$  0.026  &  1.932 $\pm$  0.365  & 17.079 $\pm$  0.022  & 29.619 $\pm$  0.684  & 17.702 $\pm$  0.033  &   9.44  &  9 \\
1996bo$^3$  & 0.017  &  1.072 $\pm$  0.010  & -0.325 $\pm$  0.113  & 15.701 $\pm$  0.005  & 24.675 $\pm$  0.186  & 16.253 $\pm$  0.013  &  85.76  & 12 \\
 1996C$^3$  & 0.030  &  1.059 $\pm$  0.038  &  0.278 $\pm$  1.085  & 16.829 $\pm$  0.052  & 27.206 $\pm$  0.928  & 17.521 $\pm$  0.030  &  23.02  & 11 \\
 1996X$^3$  & 0.007  &  1.079 $\pm$  0.041  & -1.944 $\pm$  0.371  & 13.399 $\pm$  0.012  & 24.524 $\pm$  0.774  & 14.170 $\pm$  0.033  &   7.87  & 15 \\
1997bp$^5$  & 0.008  &  1.235 $\pm$  0.049  &  1.092 $\pm$  0.277  & 14.134 $\pm$  0.006  & 25.847 $\pm$  0.484  & 14.637 $\pm$  0.018  &   8.18  & 11 \\
1997bq$^5$  & 0.009  &  1.014 $\pm$  0.014  &  0.479 $\pm$  0.099  & 14.580 $\pm$  0.017  & 25.130 $\pm$  0.230  & 15.139 $\pm$  0.017  &  16.26  & 11 \\
1997br$^5$  & 0.007  &  1.349 $\pm$  0.035  &  0.769 $\pm$  0.110  & 13.683 $\pm$  0.020  & 20.672 $\pm$  0.451  & 14.295 $\pm$  0.037  &  13.75  & 10 \\
1997cn$^5$  & 0.017  &  0.840 $\pm$  0.049  & -1.711 $\pm$  0.776  & 16.426 $\pm$  0.027  & 24.070 $\pm$  1.067  & 18.255 $\pm$  0.156  &  11.78  & 12 \\
1997dg$^5$  & 0.031  &  0.965 $\pm$  0.060  & -1.324 $\pm$  1.082  & 17.240 $\pm$  0.035  & 26.675 $\pm$  2.351  & 17.744 $\pm$  0.051  &   1.18  &  6 \\
 1997e$^5$  & 0.013  &  0.931 $\pm$  0.031  & -1.375 $\pm$  0.315  & 15.477 $\pm$  0.007  & 23.879 $\pm$  0.386  & 16.064 $\pm$  0.022  &   8.60  &  8 \\
1998ab$^5$  & 0.027  &  1.413 $\pm$  0.046  &  0.279 $\pm$  0.179  & 16.485 $\pm$  0.021  & 19.787 $\pm$  0.556  & 17.049 $\pm$  0.042  &   8.40  & 10 \\
1998dh$^5$  & 0.009  &  0.997 $\pm$  0.011  & -0.337 $\pm$  0.169  & 14.099 $\pm$  0.015  & 26.025 $\pm$  0.275  & 14.682 $\pm$  0.024  &   0.77  &  6 \\
1998es$^5$  & 0.011  &  0.980 $\pm$  0.140  & -2.309 $\pm$  0.511  & 14.083 $\pm$  0.016  & 24.278 $\pm$  1.062  & 14.928 $\pm$  0.085  &   6.90  & 11 \\
 1998v$^5$  & 0.018  &  0.940 $\pm$  0.036  &  0.545 $\pm$  0.611  & 15.790 $\pm$  0.017  & 25.688 $\pm$  0.698  & 16.100 $\pm$  0.056  &   9.65  &  7 \\
1999aa$^5$  & 0.014  &  1.288 $\pm$  0.014  &  0.301 $\pm$  0.072  & 15.242 $\pm$  0.007  & 25.585 $\pm$  0.180  & 15.871 $\pm$  0.027  &  73.43  & 14 \\
1999ac$^5$  & 0.009  &  1.210 $\pm$  0.027  &  1.328 $\pm$  0.283  & 14.321 $\pm$  0.006  & 23.045 $\pm$  0.519  & 15.057 $\pm$  0.032  &  11.26  & 12 \\
1999cl$^5$  & 0.008  &  0.970 $\pm$  0.118  & -0.107 $\pm$  0.571  & 13.139 $\pm$  0.022  & 22.206 $\pm$  1.647  & 13.688 $\pm$  0.101  &   0.63  &  8 \\
1999dq$^5$  & 0.014  &  1.179 $\pm$  0.024  & -0.324 $\pm$  0.126  & 14.785 $\pm$  0.006  & 25.492 $\pm$  0.198  & 15.312 $\pm$  0.013  &  34.19  & 20 \\
1999gp$^5$  & 0.027  &  1.261 $\pm$  0.044  & -1.600 $\pm$  0.263  & 16.405 $\pm$  0.008  & 26.011 $\pm$  0.526  & 17.045 $\pm$  0.024  &   7.51  & 11 \\
2000cn$^5$  & 0.023  &  0.799 $\pm$  0.026  & -0.362 $\pm$  0.206  & 16.679 $\pm$  0.015  & 23.965 $\pm$  0.441  & 17.256 $\pm$  0.062  &  15.40  & 12 \\
2000dk$^5$  & 0.017  &  0.809 $\pm$  0.016  & -1.342 $\pm$  0.177  & 15.767 $\pm$  0.007  & 24.297 $\pm$  0.429  & 16.222 $\pm$  0.048  &  20.33  &  9 \\
2000fa$^5$  & 0.021  &  1.119 $\pm$  0.050  & -0.235 $\pm$  0.278  & 16.289 $\pm$  0.037  & 24.911 $\pm$  0.423  & 16.888 $\pm$  0.111  &   0.20  &  7 \\
\hline
\end{tabular}
\caption{Results of the $I$-band light curve fit of 42 nearby
  supernovae: $t_1$ and $I_1$ are the parameters for the time and
  amplitude fitted on the first template, $t_2$ and $I_2$ are the
  parameters for the time and amplitude fitted on the second template,
  and $s_I$ is the stretch factor. $N$ is the number of points used in
  the fit ($dof$=$N$-5). The data were taken from: $^1$~\citet{89b};
  $^2$~\citet{94d}; $^3$~\citet{Riess22}; $^4$~\citet{Hamuy1996};
  $^5$~\citet{jha}; $^6$~\citet{Filippenko92,leibundgut93}. }
\label{Ifit_results}
\end{center}
\end{table*}

\begin{figure}[htb]
\centering \includegraphics[width=8cm]{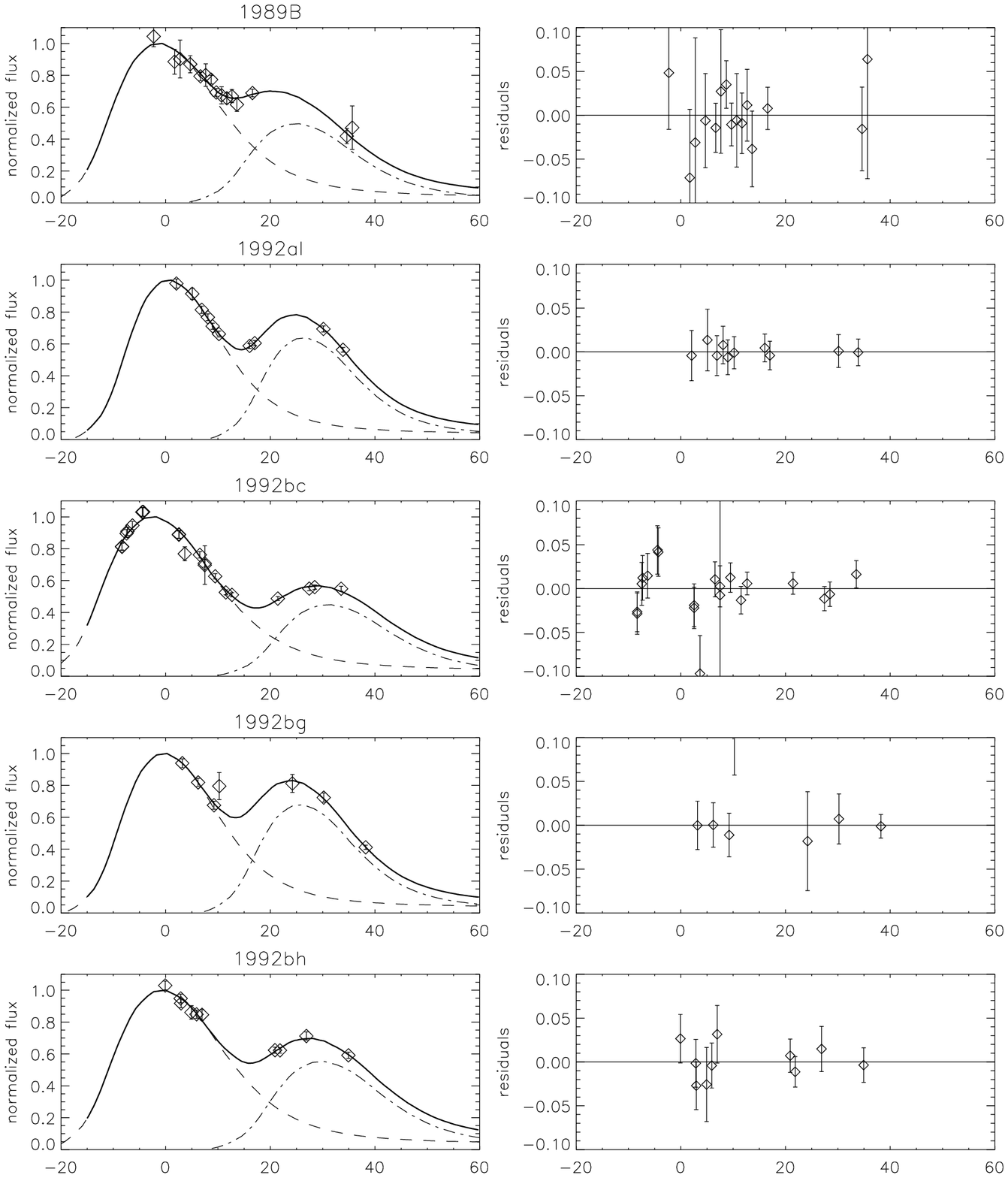}
\centering \includegraphics[width=8cm]{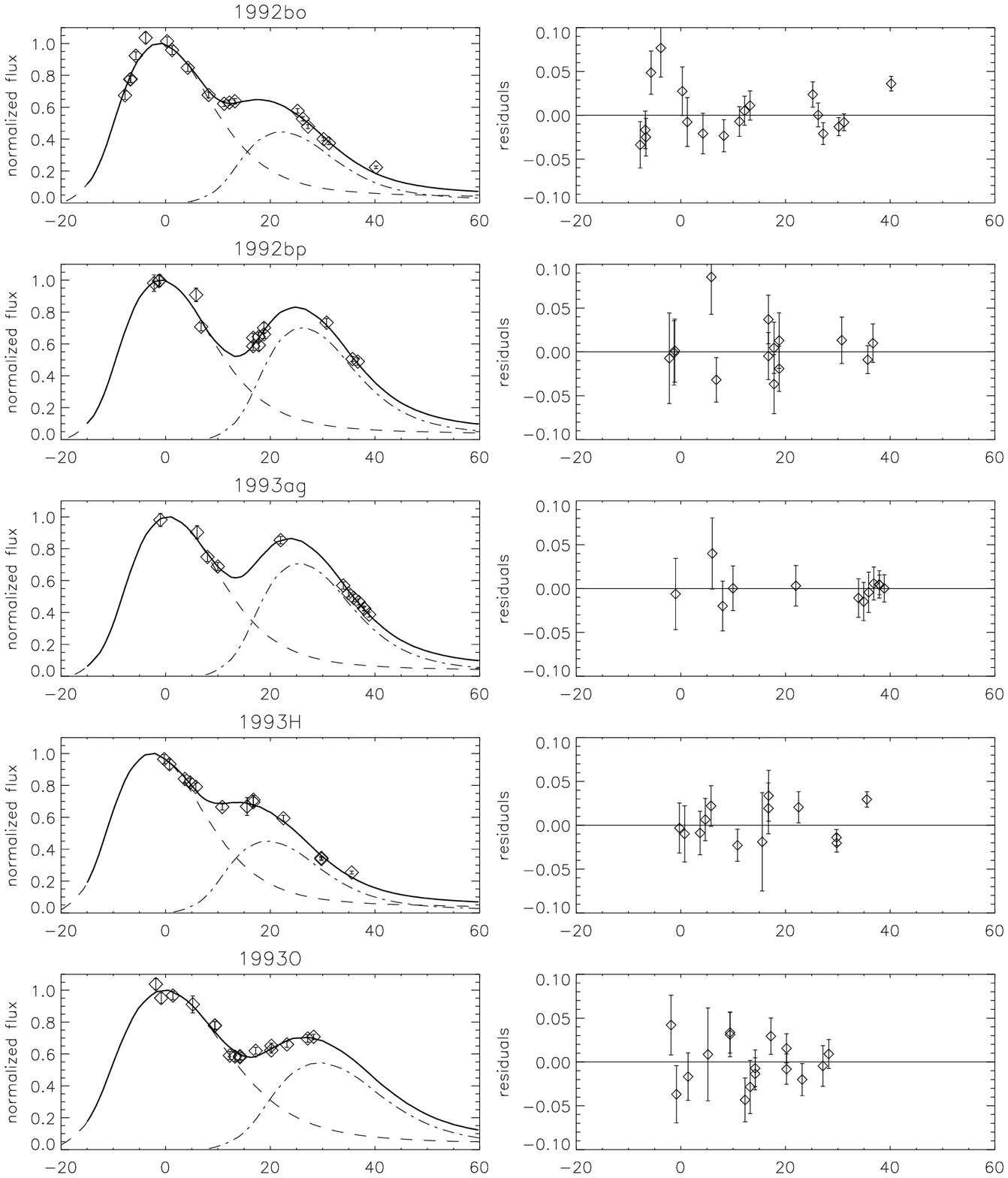}
\caption{$I$-band light curve fits. On the ordinate is the flux
  normalised to the first peak, on the abscissa the restframe time
  since $B$-band maximum. The dashed line and the dash-dotted line
  represent the two templates used to fit the first and
  second peak respectively.}
\label{plot9}
\end{figure}

\setcounter{figure}{0}
\begin{figure}[htb]
\centering \includegraphics[width=8cm]{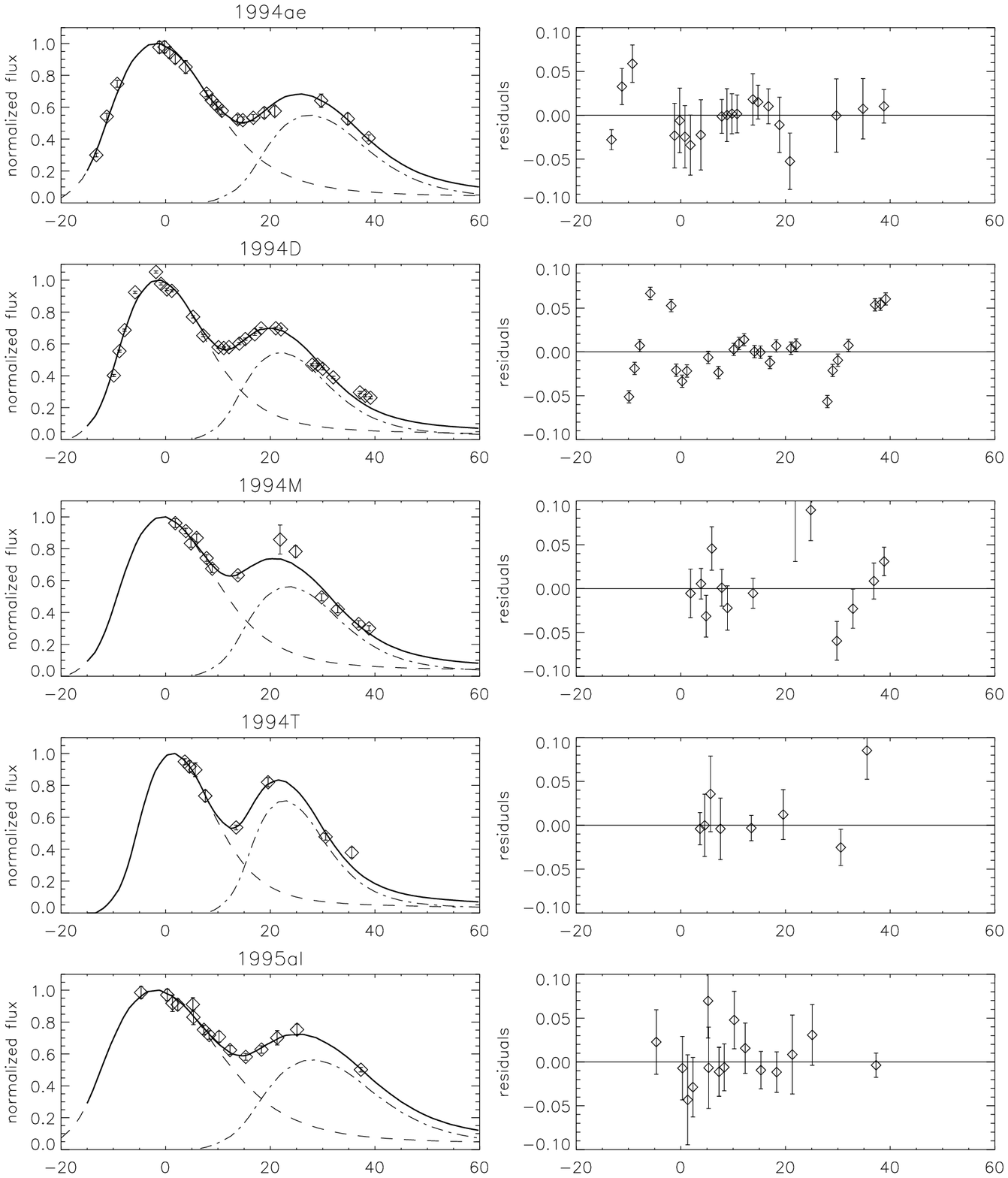}
\centering \includegraphics[width=8cm]{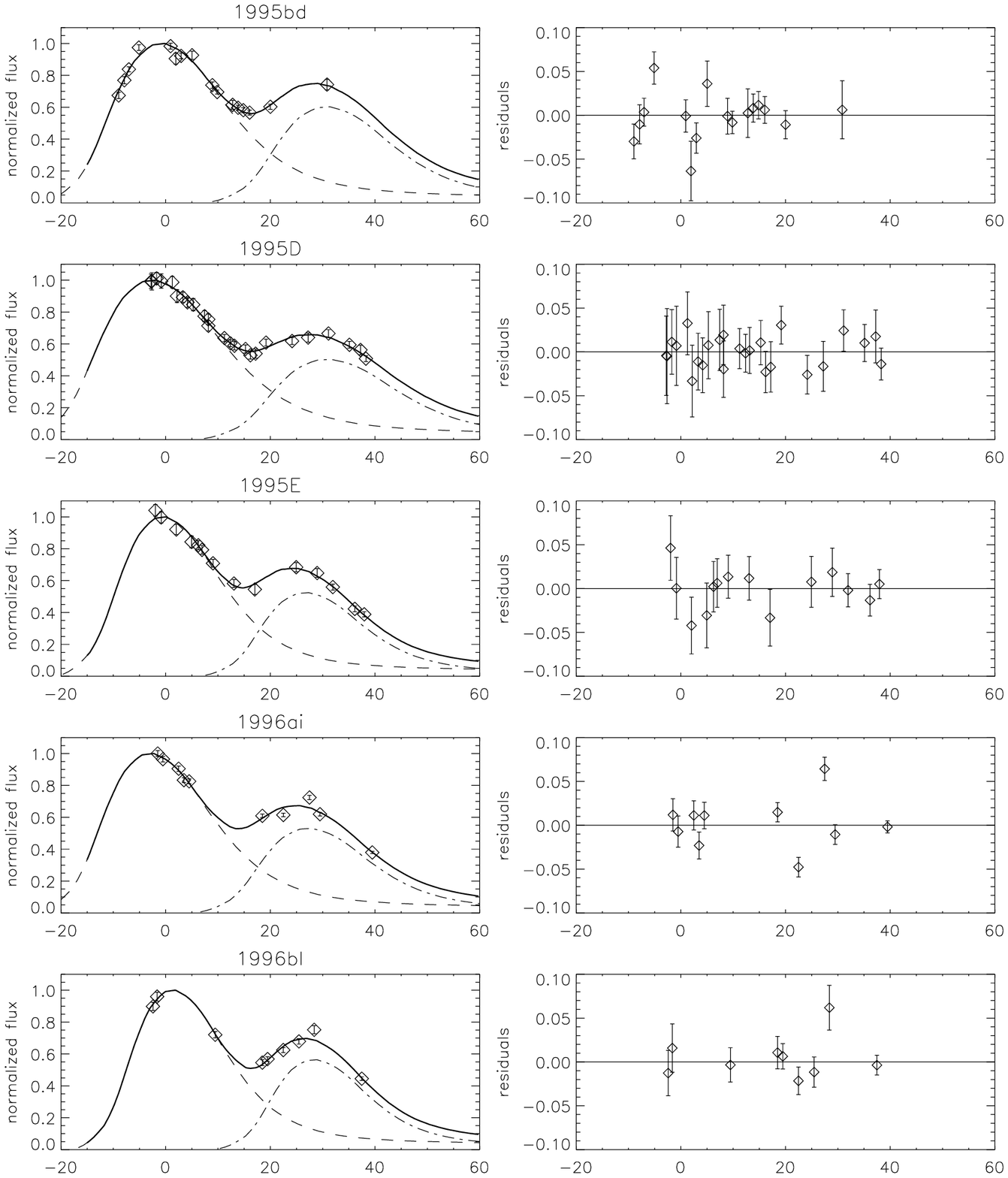}
\caption{continued. $I$-band light curve fits. On the ordinate is the
  the flux normalised to the first peak, on the abscissa the restframe
  time since $B$-band maximum. The dashed line and the dash-dotted
  line represent the two templates used to fit the first and
  second peak respectively.}
\label{plot1}
\end{figure}

\setcounter{figure}{0}
\begin{figure}[htb]
\centering \includegraphics[width=8cm]{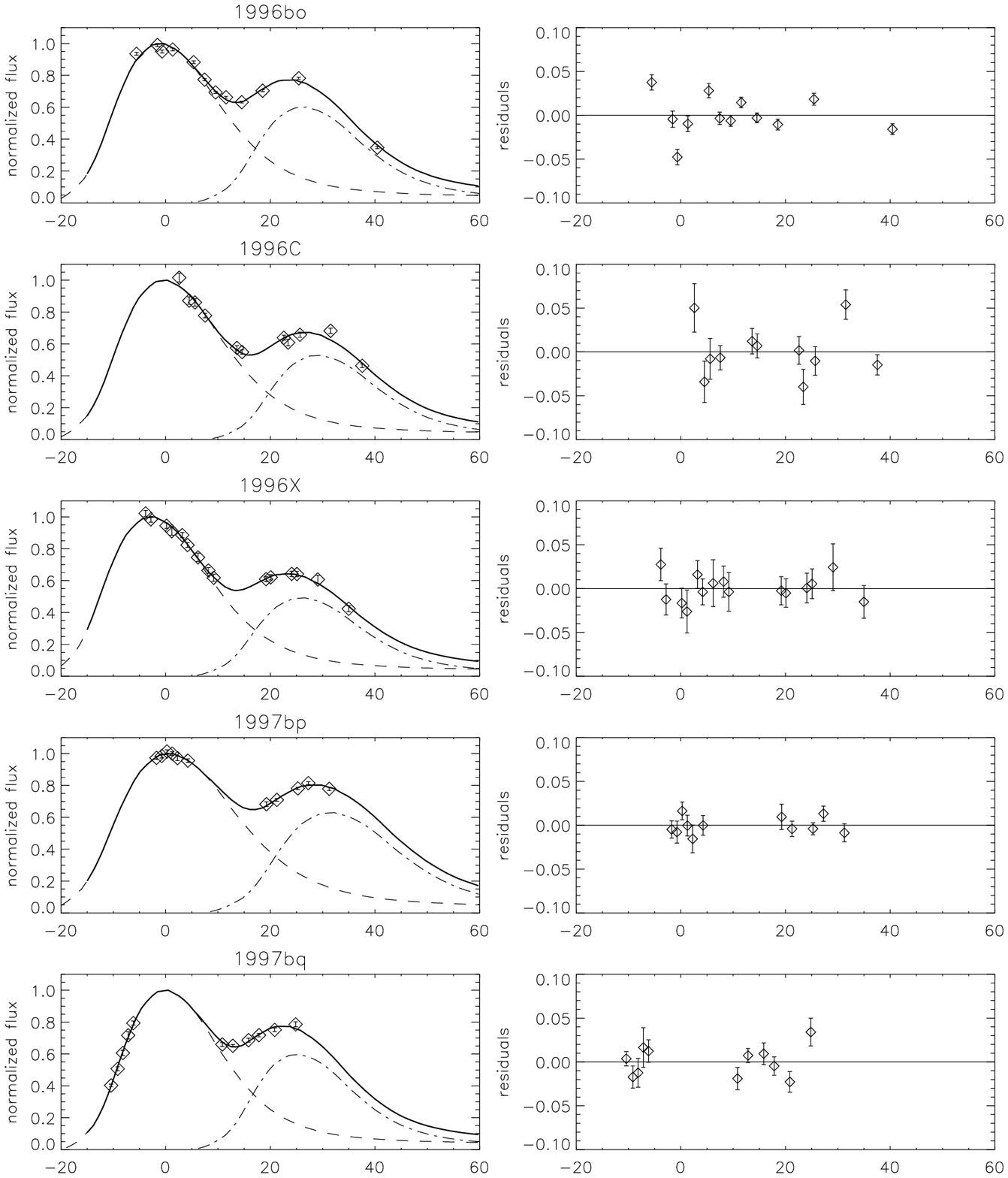}
\centering \includegraphics[width=8cm]{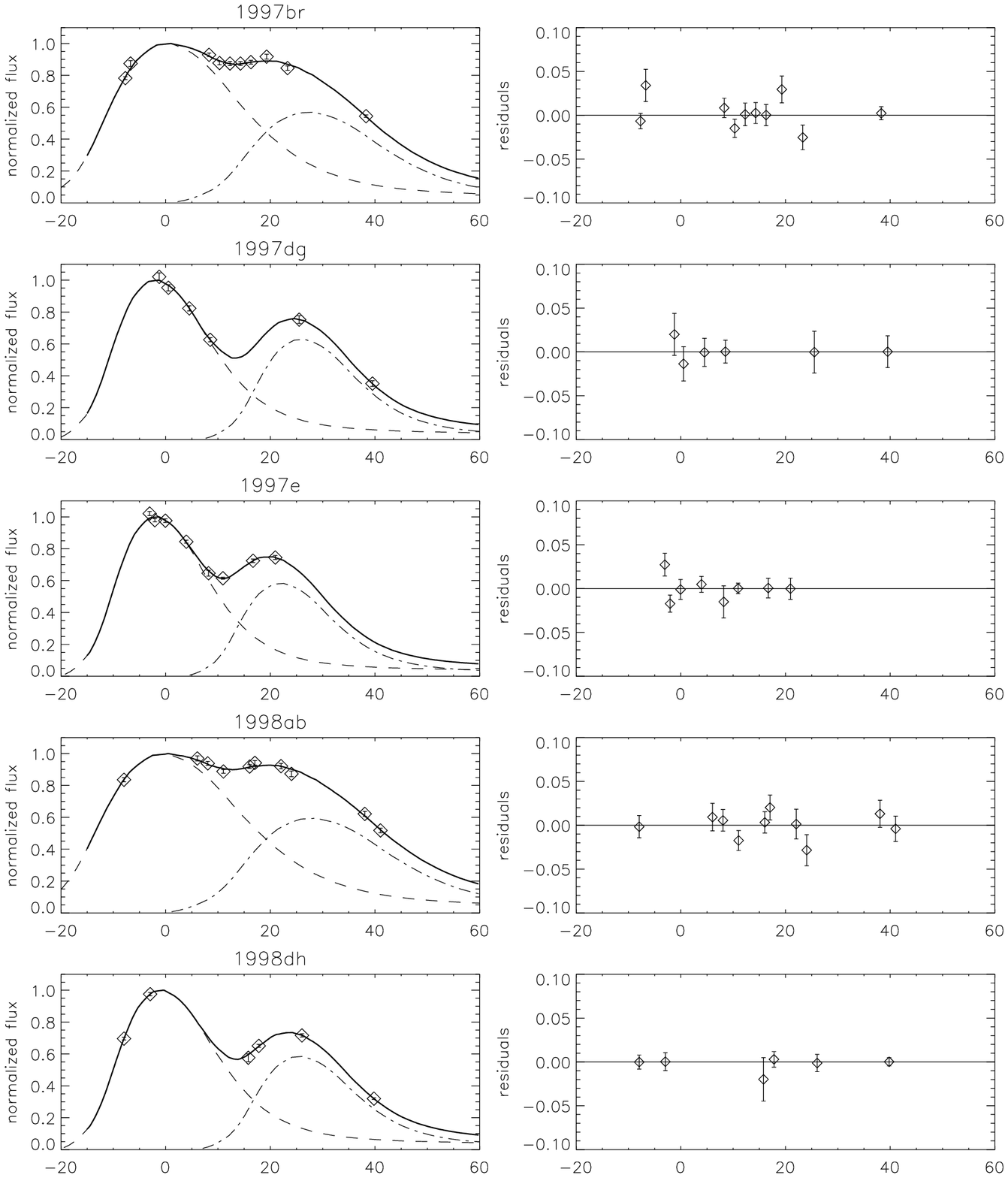}
\caption{continued. $I$-band light curve fits. On the ordinate is the
  flux normalised to the first peak, on the abscissa the restframe
  time since $B$-band maximum. The dashed line and the dash-dotted
  line represent the two templates used to fit the first and
  second peak respectively.}
\label{plot3}
\end{figure}

\setcounter{figure}{0}
\begin{figure}[htb]
\centering \includegraphics[width=8cm]{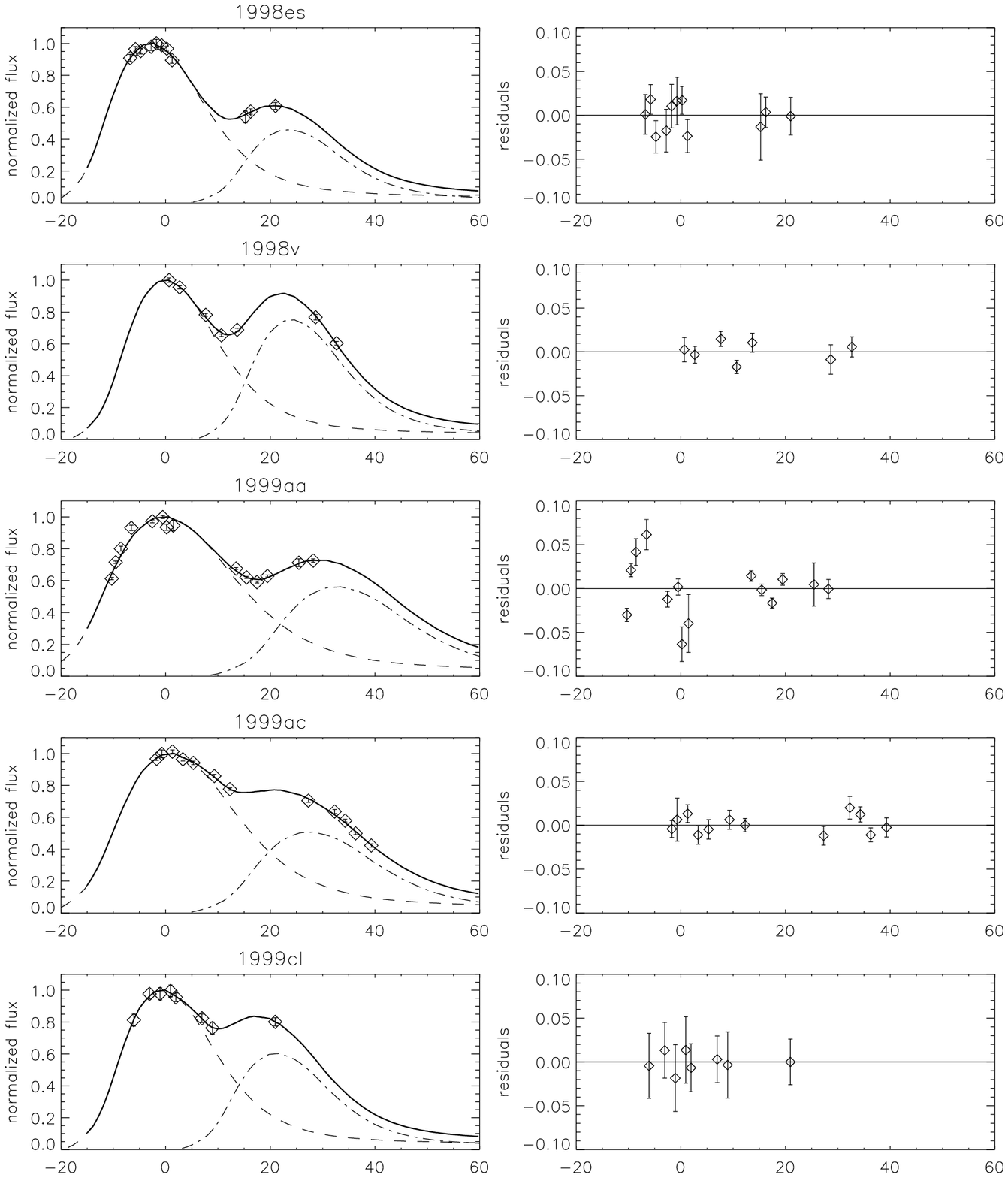}
\centering \includegraphics[width=8cm]{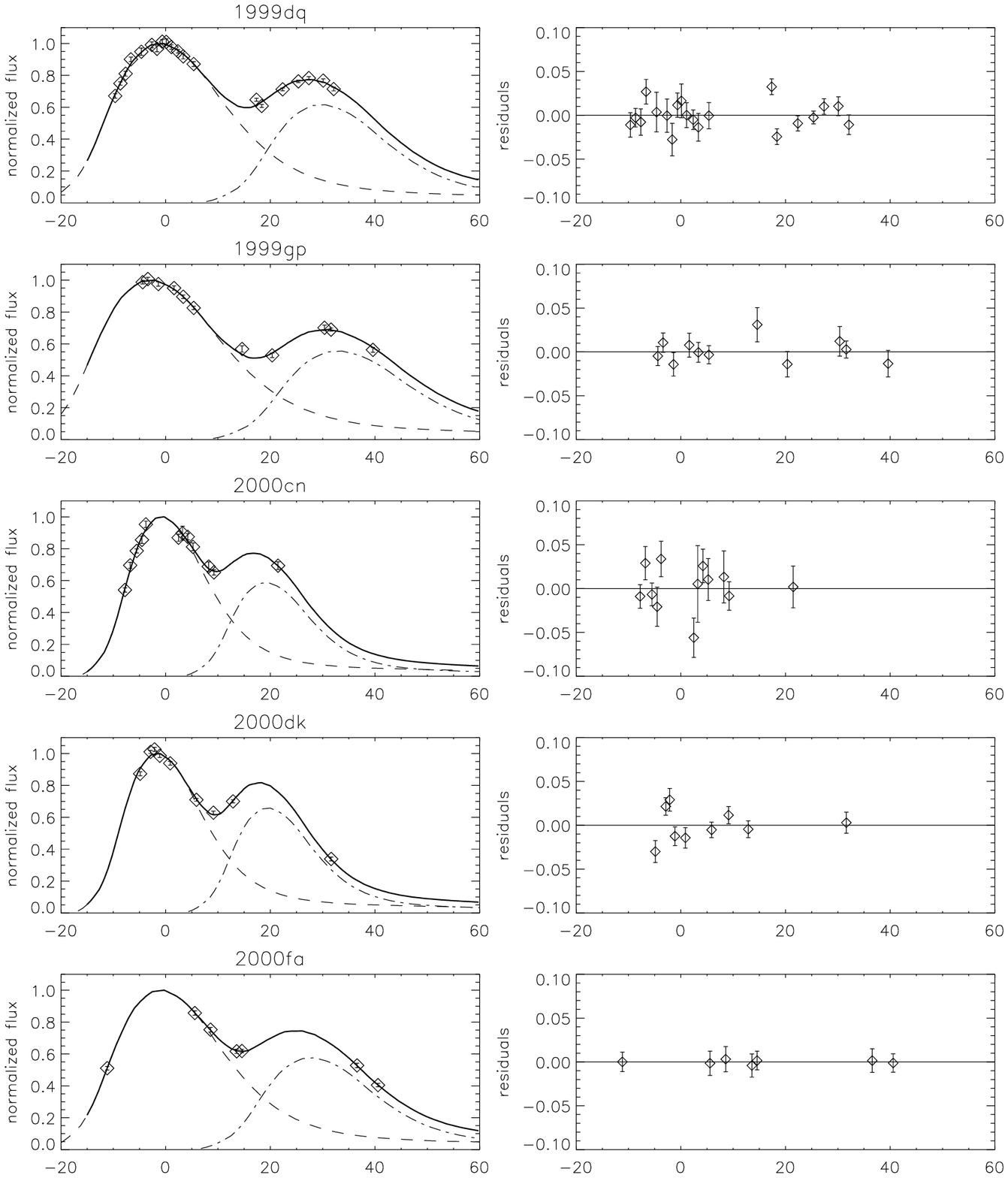}
\caption{continued. $I$-band light curve fits. On the ordinate is the
  flux normalised to the first peak, on the abscissa the restframe
  time since $B$-band maximum. The dashed line and the dash-dotted
  line represent the two templates used to fit the first and
  second peak respectively.}
\label{plot5}
\end{figure}

\setcounter{figure}{0}
\begin{figure}[htb]
\centering \includegraphics[width=8cm]{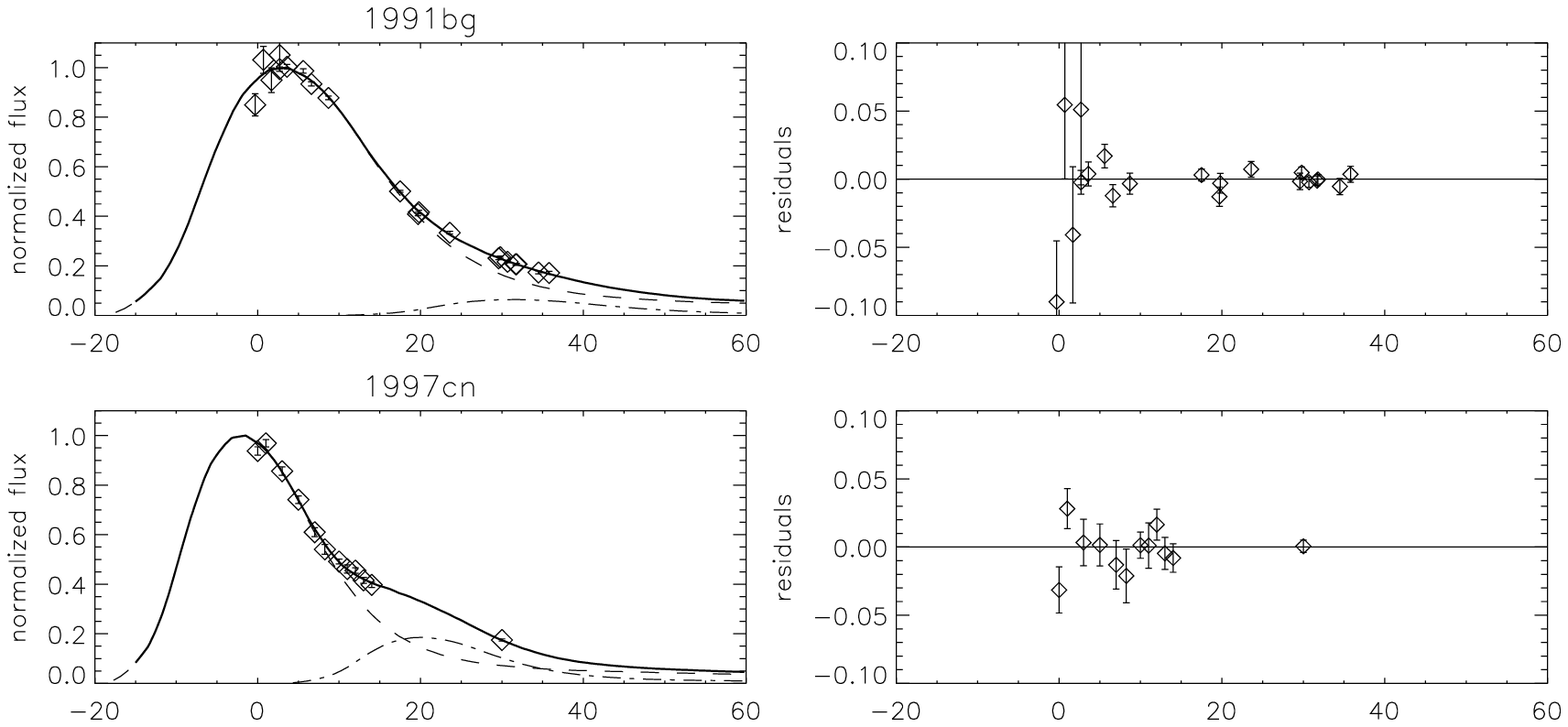}
\caption{continued. $I$-band light curve fits of the underluminous supernovae
 SN~1991bg and SN~1997cn. The dashed line and the dash-dotted line
 represent the two templates used to fit the first and second
 peak respectively. Note that the second peak is $\sim 3$
 mag fainter than the first peak.}
\label{subluminous}
\end{figure}

\begin{figure}[htb]
\centering \includegraphics[width=4cm]{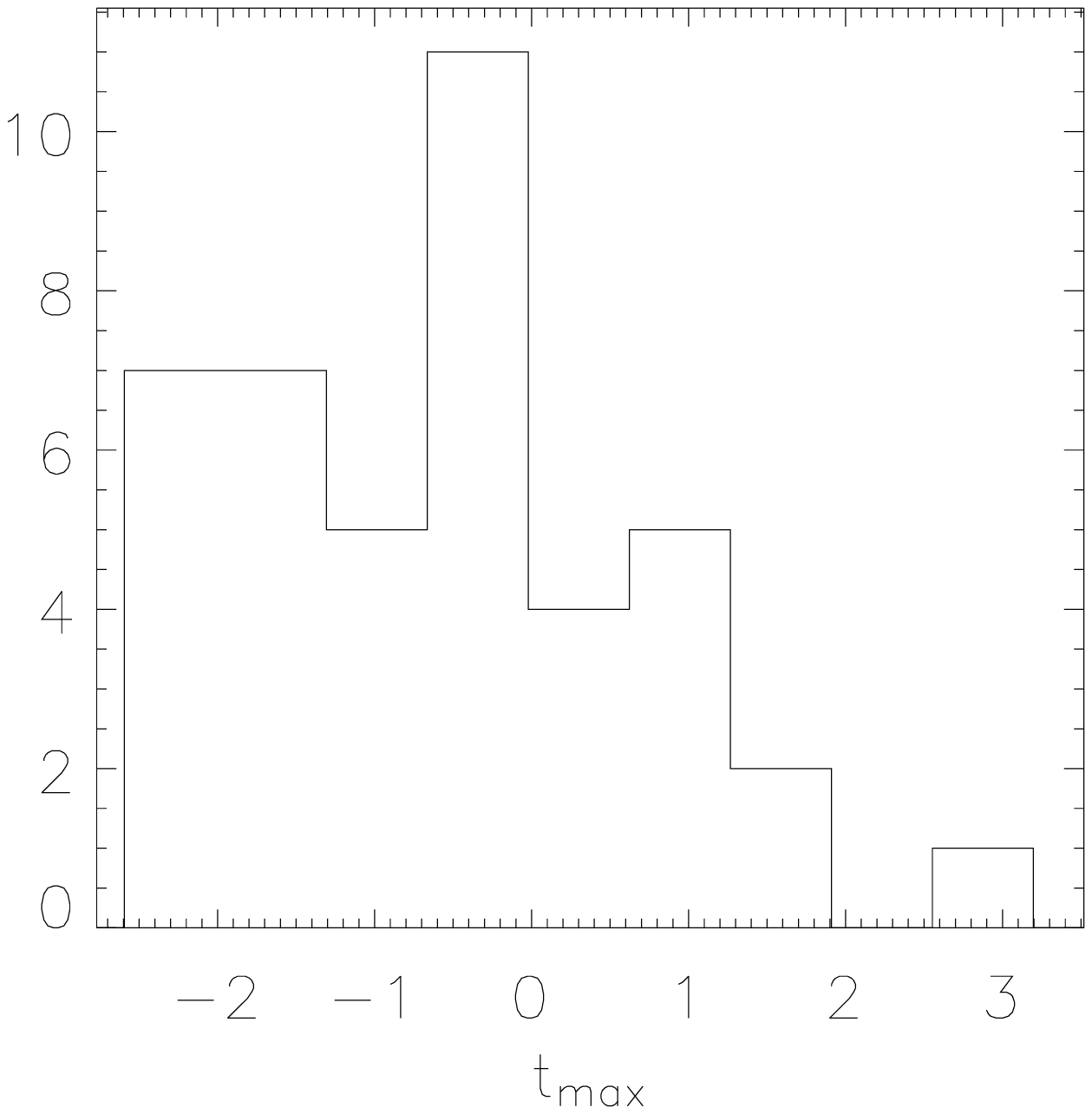}
\centering \includegraphics[width=4cm]{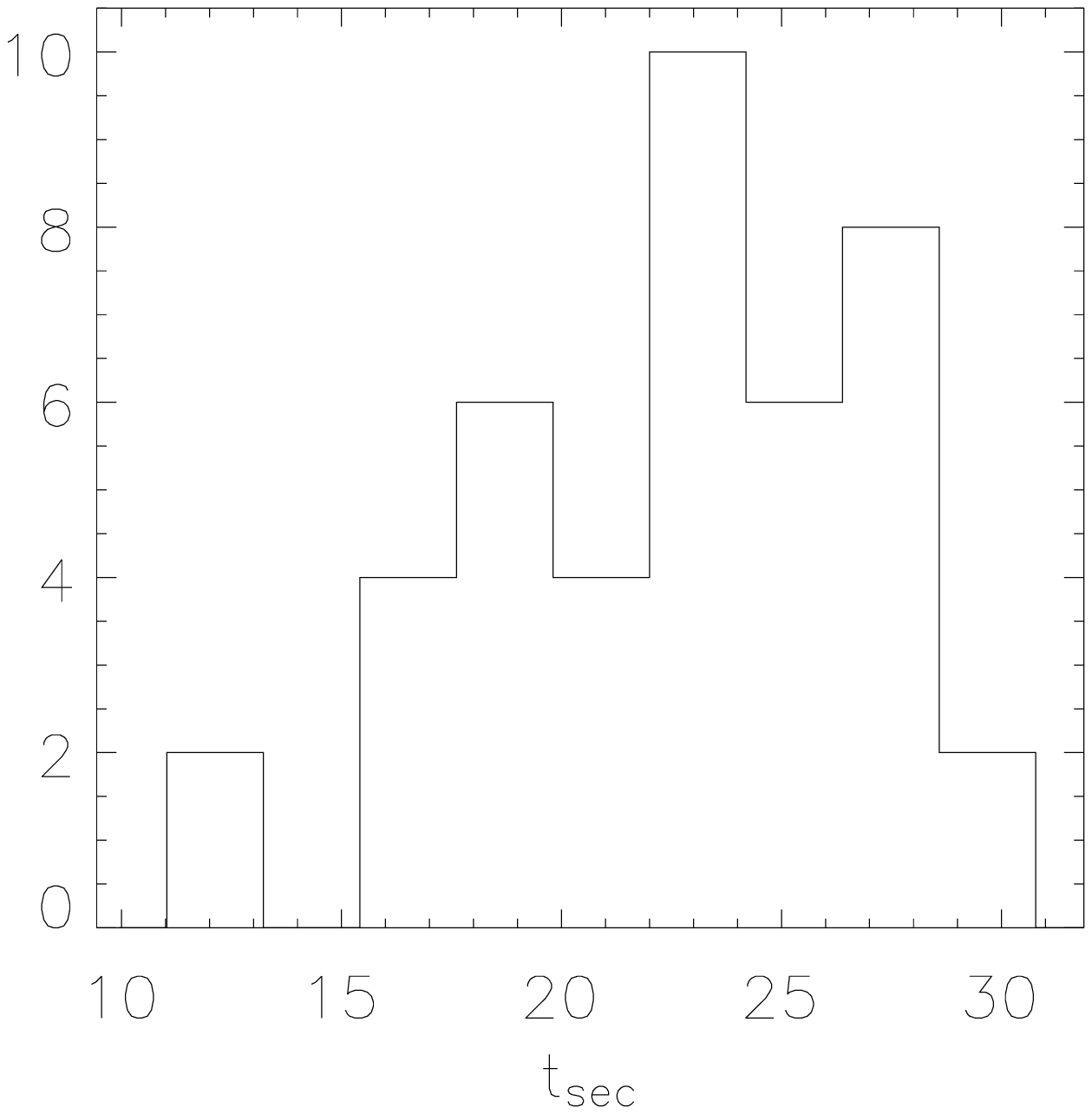}
\caption{Distribution of the time of $I$-band maximum referred to the
  time of $B$-band maximum (left panel) and the distribution of the
  time of second maximum referred to the time of $B$-band maximum
  (right panel).}
\label{histo_params}
\end{figure}

\subsection{Monte-Carlo tests of the fitting method}

Given the heterogeneous origin of the data sample, the quality and the
sampling of the individual SN light curves vary considerably. Only a
few supernovae have excellent time coverage in the $I$-band, resulting
in a wide range of accuracy in the fitted parameters. The robustness
of the fitting procedure was tested for all circumstances of data
quality and time sampling in our sample by means of Monte Carlo
simulations. We generated 1000 sets of simulated light curves for each
supernova. The synthetic data points had the same time sampling as the
real light curves and with deviations from the best fit template
randomly drawn from a Gaussian whose width was set by the published
uncertainties. The simulated light curves were fitted using the same
method as the experimental data sets. The distribution of the
fitted parameters from the simulated data was compared with the input
data from the fits of the experimental data. The mean value in the
distribution of each parameter generally coincides with that expected,
i.e. within one standard deviation. There is no evidence for biased
fit parameters. This lends confidence that the fitting procedure is
robust, and given the model of the light curve template, will not
yield biased estimation of the parameters. 

In two cases, SN~1997br and SN~1998ab, we found that the fits to the
MC simulations resulted in two solutions, one corresponding to that
found in the fit to the real data and the other corresponding to a
small fraction (3\% and 22\% for SN~1997br and SN~1998ab respectively)
of all simulations. We note, incidentally, that these SNe are the two
with the smallest ratios between peak and dip in their light
curves. However, the first peak is determined by only one and two data
points each.  A close look at the simulated light curves indicates the
limited number of points constraining the peak is the cause of the
rare failure of the MC simulation. We nevertheless keep these SNe in
the rest of the analysis, since the parameters and their uncertainties
estimated from the main distribution agree with the results on the
real data.

For the rest of the supernovae, the simulations confirmed the expected
parameters, giving general confidence in the robustness of the
procedure and the accuracy of the uncertainties on the parameters
given in Table~\ref{Ifit_results}.

\subsection{Intrinsic variations}

\begin{figure}[htb]
\centering \includegraphics[width=8cm]{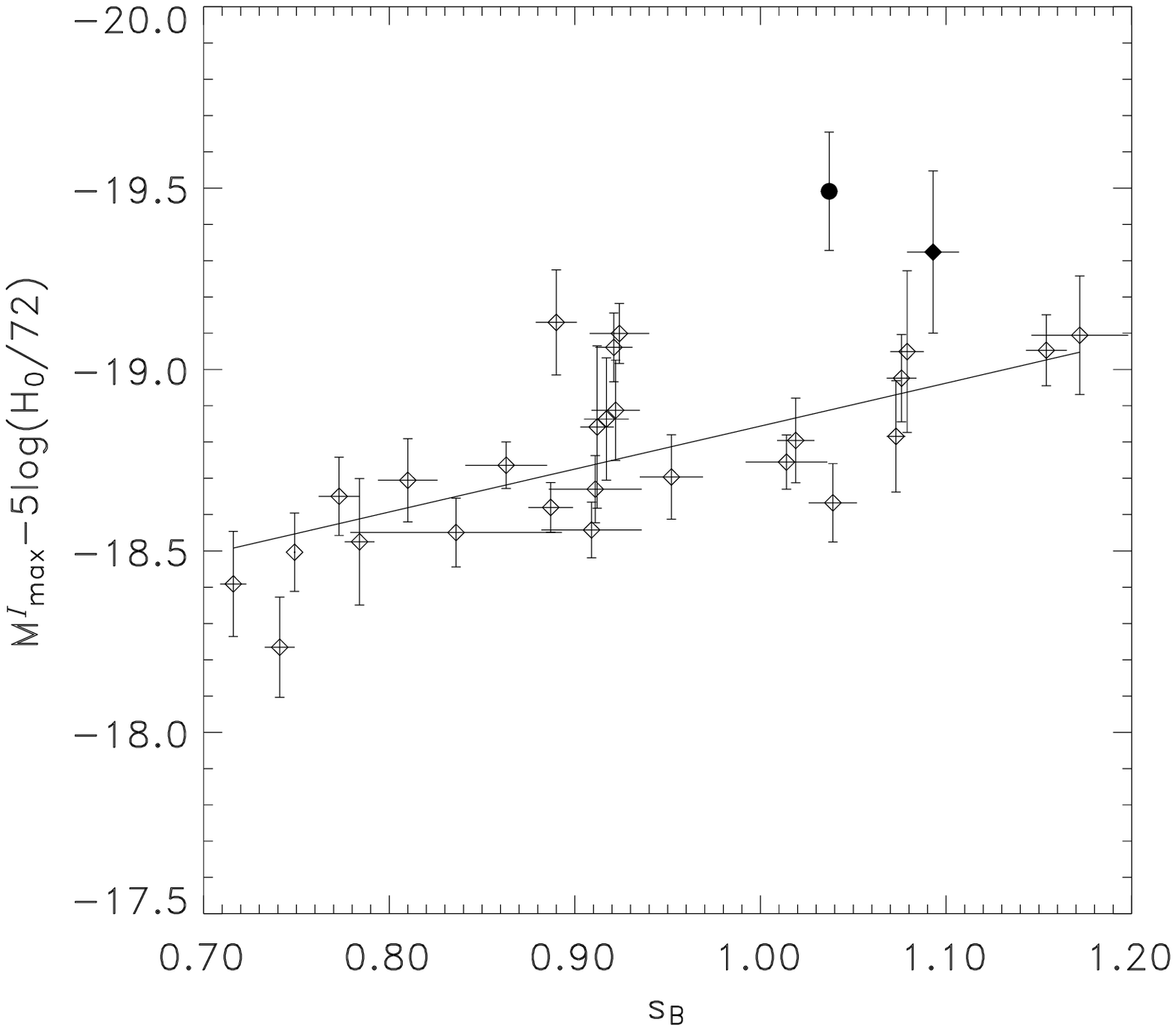}
\caption{$I$-band absolute magnitude versus stretch in the $B$-band. The
  best fit gives $\alpha_I = 1.18 \pm 0.19$ and $M_{\rm
  max}^I(s_B=1)=-18.84 \pm 0.03$ mag. The two deviating supernovae,
  SN~1998es (filled diamond) and SN~1999dq (filled circle), were
  excluded from the fit.}
\label{mag_stretch.b}
\end{figure}

We investigated possible relations between $I$-band and $B$-band
parameters. Following \citet{gerson}, the time of maximum, the stretch
factor, $s_B$, and the amplitude of maximum, $m_B$, were determined by
fitting a $B$-band template to the published $B$-band data.  A
width-luminosity relation was found for the first $I$-band light curve
peak. Figure \ref{mag_stretch.b} shows the $I$-band absolute magnitude
versus the stretch factor in the $B$-band for SNe with $z_{CMB} \ge
0.01$, where the distance (in Mpc) to each SN was calculated from its
redshift, assuming a value for the Hubble constant, $H_0=72$ km
s$^{-1}$ Mpc$^{-1}$. The error bars in Fig.~\ref{mag_stretch.b}
include an uncertainty of $300$ km s$^{-1}$ on the redshifts to
account for the peculiar velocities of the host galaxies. The
underluminous supernovae, SN~1997cn and SN~1991bg, are not included in
the sample or in any of the analysis presented in this section.
Corrections for Milky Way and host galaxy extinction were also
applied, i.e.
$$M_{\rm max}^I-5\log(H_0/72)=I_{\rm max}-A_I^{MW}-A_I^{host}-25-5 \log(d_L)$$

\noindent The host galaxy extinction correction that is applied to
most of the supernovae is the weighted average of the three estimates
given in Table 2 of \citet{Phillips} assuming $R_I=1.82$.  The
extinction for the supernovae in the CfA2 data set was calculated
following the same procedure, using $B$ and $V$-band photometry. At
this point, we exclude SN~1995E, SN~1996ai and SN~1999cl from the
sample as they are highly reddened (see also discussion in
\citet{nobili2003}).  These are not shown in any of the plots nor used
in any of the analysis that follow.  Two supernovae in the sample, the
spectroscopically peculiar SN~1998es and SN~1999dq, plotted with
filled symbols in Fig.~\ref{mag_stretch.b}, appear intrinsically
redder than average, and become $\sim 2-3\sigma$ deviant from the
average after correction for host galaxy extinction. Before
introducing light curve shape corrections, the spread measured in
$M_{\rm max}^I$ excluding these two SNe, is about 0.24 mag (0.28 mag
if they are included). The solid line shows the best fit to the data,
obtained for a slope $\alpha_I = 1.18\pm 0.19$ and an absolute
magnitude for a stretch $s_B=1$ supernova equal to $M_{\rm
max}^I(s_B=1)=-18.89 \pm 0.03$ mag~\footnote{The value fitted for
$M^I_{\rm max}$ depends on the value assumed for the Hubble parameter,
$H_0=72$ km s$^{-1}$ Mpc$^{-1}$. However, its value is not used in any
of the further analysis presented in this paper.}. The dispersion,
computed as the r.m.s.  about the fitted line is 0.17 $\pm$ 0.03 mag.
A similar correlation was found between the peak magnitude and the
stretch in the $I$-band, $s_I$, with a r.m.s. of $\sim$~0.19~mag about
the best fit line, again excluding SN~1998es and SN~1999dq.

\begin{figure}[htb]
\centering \includegraphics[width=7cm]{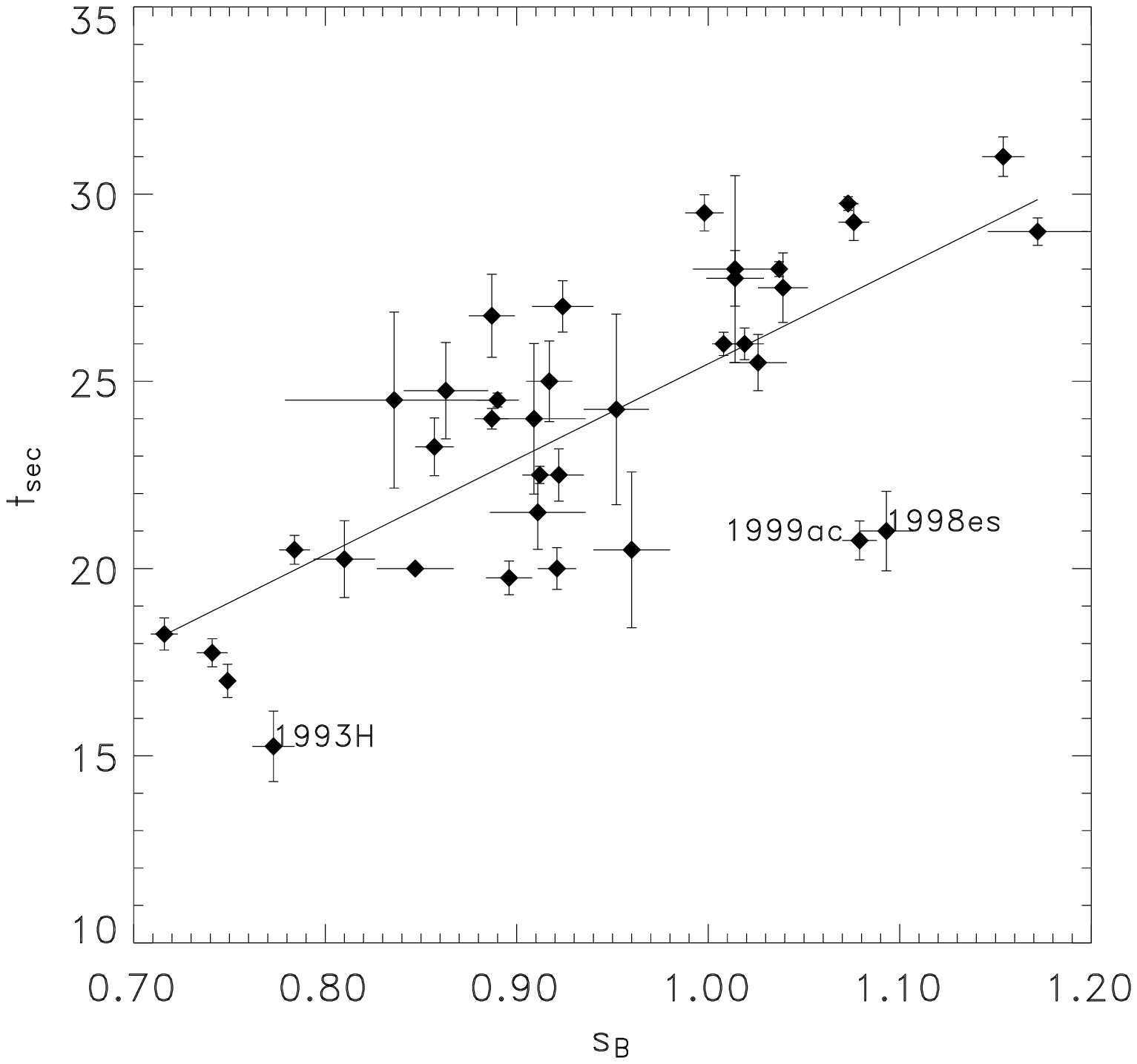}
\caption{Time since $B_{\rm max}$ of the second peak versus the stretch in
  the $B$-band. The labelled supernovae, classified as spectroscopically
  peculiar, are excluded from the fit because they are outliers.}
\label{t2_stretch}
\end{figure}

A correlation was found between $t_{\rm sec}$ and the $B$-band stretch
factor, as shown in Fig.~\ref{t2_stretch}. There are three outliers
labelled in the figure, SN~1993H , SN~1998es and SN~1999ac, which are
identified as spectroscopically peculiar supernovae.  However, other
supernovae in our sample that are classified as spectroscopically peculiar
behave as ``normal'' Type~Ia SNe.  We note that the $B$-band stretch
factor for SN~1999ac is not well defined due to an asymmetry of the
$B$-band light curve \citep{phillips99ac}.  

\begin{figure}[htb]
\centering 
\includegraphics[width=7cm]{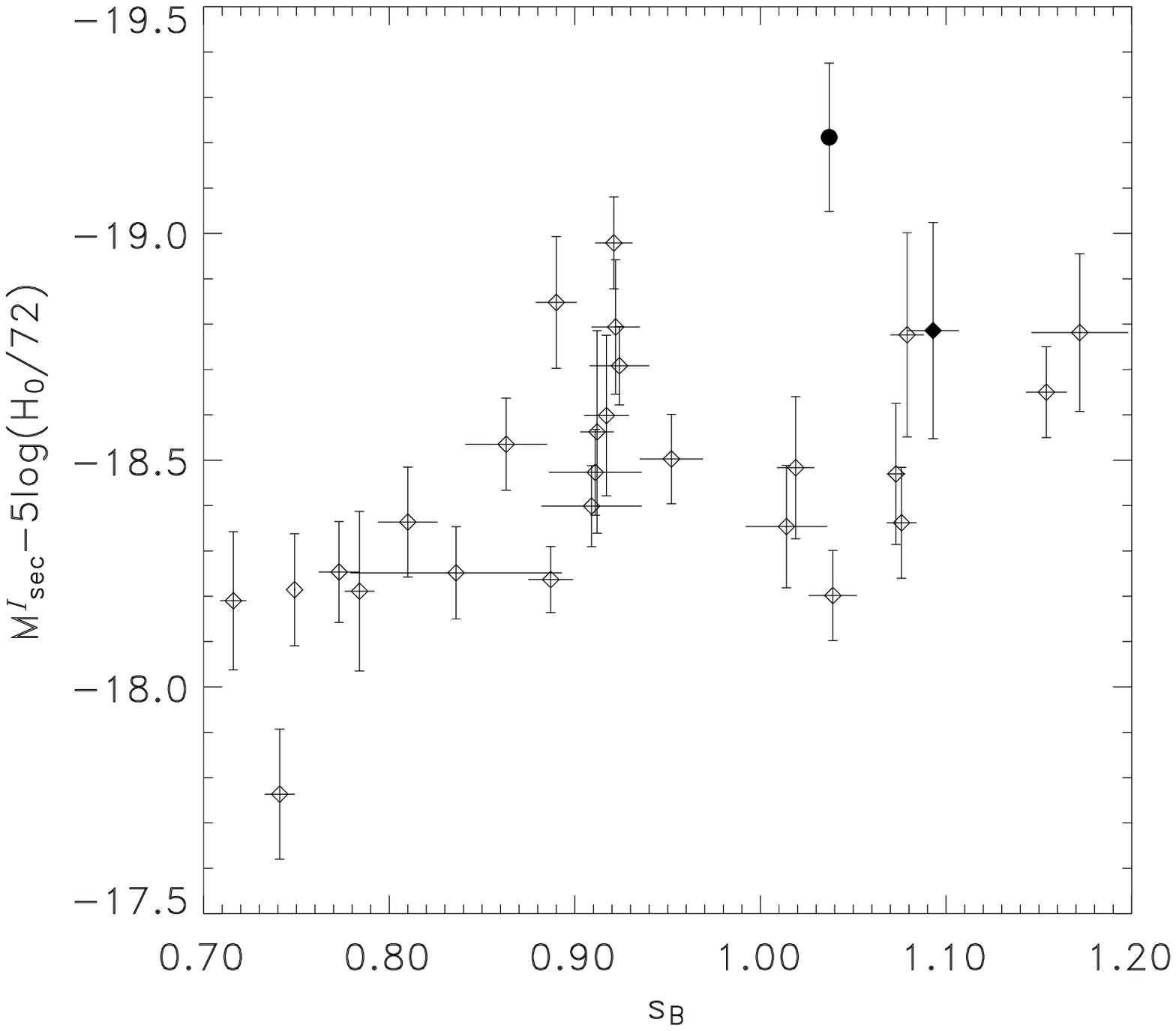}
\caption{Absolute magnitude of the secondary peak versus the stretch
  in the $B$-band. The two deviating supernovae in
  Fig.~\ref{mag_stretch.b} are SN~1998es (filled diamond) and
  SN~1999dq (filled circle).}
\label{I2_stretch}
\end{figure}

Figure~\ref{I2_stretch} shows a possible correlation between $I_{\rm
sec}$ and the stretch $s_B$, at least for $s_B<0.9$, after correcting
for the luminosity distance and for extinction both from host galaxy
and Milky Way. This correlation, however, disappears for larger values
of $s_B$.

All of these correlations, shown in Fig.~\ref{mag_stretch.b} -
~\ref{I2_stretch}, were expected since it has been suggested that the
location and the intensity of the secondary peak depends on the
$B$-band intrinsic luminosity of the supernova \citep{Hamuy1996}.

\begin{figure}[htb]
\centering \includegraphics[width=7cm]{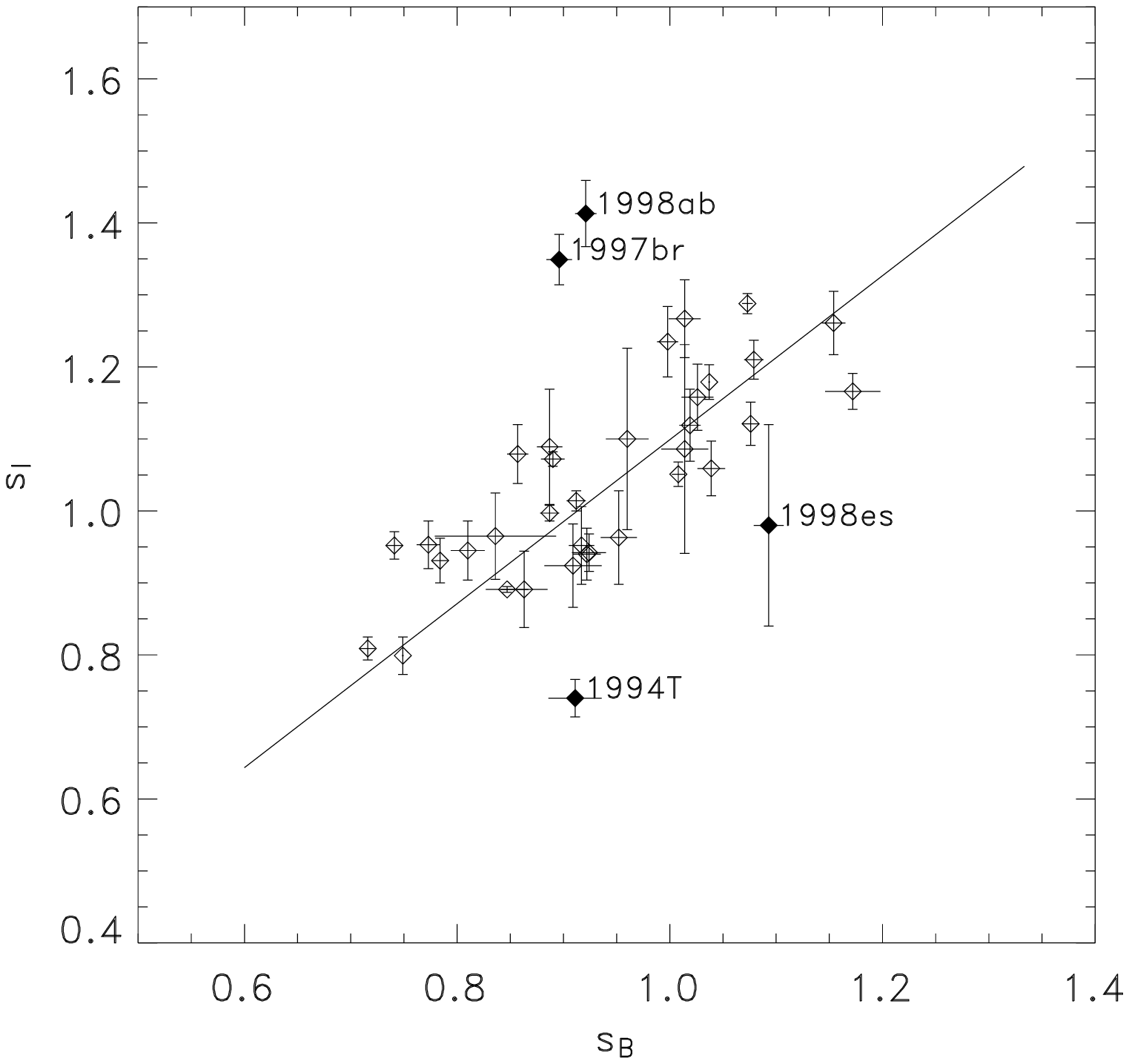}
\caption{ $I$-band light curve stretch, $s_I$, versus $B$-band
stretch, $s_B$. The labelled supernovae are more than two standard
deviation from the correlation shown by the ensemble. SN~1998ab,
SN~1997br and SN~1998es are classified as spectroscopically peculiar.}
\label{si_sb}
\end{figure}

Figure~\ref{si_sb} shows the $I$-band stretch, $s_I$, plotted versus
the $B$-band stretch, $s_B$. We found an interesting linear
correlation, although some of the supernovae, three of which are
spectroscopically peculiar, are more than two standard deviations from
the fit. The dispersion measured as r.m.s. about the line is 0.08.

We have investigated the possible existence of further relations
between the fitted parameters, but find no additional statistically
significant correlations.

\begin{table}[htb]
\begin{center}
\begin{tabular}{cccc}\hline
SN &  $s_B$ &  $z_{CMB}$  & $m_I^{\rm eff}$\\
\hline\hline
  1992al  &  0.917 $\pm$  0.012  &  0.014  &  14.868 $\pm$  0.162  \\
  1992bc  &  1.076 $\pm$  0.008  &  0.020  &  15.718 $\pm$  0.111  \\
  1992bg  &  0.952 $\pm$  0.017  &  0.036  &  17.120 $\pm$  0.107  \\
  1992bh  &  1.014 $\pm$  0.022  &  0.045  &  17.637 $\pm$  0.062  \\
  1992bo  &  0.741 $\pm$  0.008  &  0.017  &  15.710 $\pm$  0.138  \\
  1992bp  &  0.863 $\pm$  0.022  &  0.079  &  18.689 $\pm$  0.055  \\
   1993H  &  0.773 $\pm$  0.011  &  0.025  &  16.169 $\pm$  0.106  \\
   1993O  &  0.887 $\pm$  0.012  &  0.053  &  17.967 $\pm$  0.054  \\
  1993ag  &  0.909 $\pm$  0.027  &  0.050  &  17.928 $\pm$  0.069  \\
   1994M  &  0.810 $\pm$  0.016  &  0.024  &  16.080 $\pm$  0.111  \\
   1994T  &  0.911 $\pm$  0.025  &  0.036  &  17.105 $\pm$  0.085  \\
  1995bd  &  1.172 $\pm$  0.026  &  0.014  &  14.939 $\pm$  0.162  \\
   1996C  &  1.039 $\pm$  0.013  &  0.027  &  16.669 $\pm$  0.097  \\
  1996bl  &  0.924 $\pm$  0.016  &  0.035  &  16.630 $\pm$  0.070  \\
  1996bo  &  0.890 $\pm$  0.011  &  0.016  &  14.860 $\pm$  0.138  \\
  1997bq  &  0.912 $\pm$  0.009  &  0.010  &  14.154 $\pm$  0.219  \\
  1997dg  &  0.836 $\pm$  0.057  &  0.030  &  16.740 $\pm$  0.110  \\
   1997E  &  0.784 $\pm$  0.008  &  0.013  &  14.888 $\pm$  0.173  \\
  1998ab  &  0.921 $\pm$  0.010  &  0.028  &  16.180 $\pm$  0.083  \\
  1998es  &  1.093 $\pm$  0.014  &  0.010  &  13.885 $\pm$  0.219  \\
   1998V  &  0.922 $\pm$  0.013  &  0.017  &  15.272 $\pm$  0.131  \\
  1999aa  &  1.073 $\pm$  0.005  &  0.015  &  15.250 $\pm$  0.146  \\
  1999ac  &  1.079 $\pm$  0.009  &  0.010  &  14.144 $\pm$  0.218  \\
  1999dq  &  1.037 $\pm$  0.000  &  0.014  &  14.382 $\pm$  0.155  \\
  1999gp  &  1.154 $\pm$  0.011  &  0.026  &  16.303 $\pm$  0.090  \\
  2000cn  &  0.749 $\pm$  0.000  &  0.023  &  16.114 $\pm$  0.107  \\
  2000dk  &  0.716 $\pm$  0.007  &  0.016  &  15.375 $\pm$  0.147  \\
  2000fa  &  1.019 $\pm$  0.010  &  0.022  &  16.029 $\pm$  0.106  \\
\hline
\end{tabular}
\caption{List of SNe used in the Hubble diagram. $m_I^{\rm eff}$ is
  the peak magnitude corrected for dust extinction and for the
  width-luminosity relation, following Eq.~\ref{Imax}.  The quoted
  uncertainties do not include the redshift contribution due to
  peculiar velocities in the host galaxies, assumed equal to 300 km
  s$^{-1}$. Redshifts from Table~\ref{Ifit_results} are here
  transformed into the CMB frame.}
\label{Iband}
\end{center}
\end{table}

\section{The $I$-band Hubble Diagram}
\label{sec:hd}

The fitted values of $I_{\rm max}$ were used to build a Hubble diagram
in the $I$-band. We select 28 supernovae from the sample considered
here that have a redshift $z_{CMB} \ge 0.01$\footnote{The lower limit
chosen in previous analyses by the SCP is slightly higher. However, we
include these lower redshift SNe in the sample in order to increase
the statistical significance. Cutting the Hubble diagram above $z$ =
0.015 would decrease the sample by about 30 \%. Note, however, that
this choice does not significantly affect any of the results.}. The
maximum redshift in this sample is 0.1.

The width-luminosity relation between $I_{max}$ and the the $B$-band
stretch factor was used to correct the peak magnitude, with a
$\alpha_I = 1.18 \pm 0.19$ as measured in the previous section,
similarly to what is usually done in the $B$-band
\citep{perlmutter42}. The peak magnitude was also corrected for Milky
Way and host galaxy extinction:

\begin{equation}
m_I^{\rm eff}=m_I+\alpha_I(s_B-1)-A_I^{host}-A_I^{MW}
\label{Imax}
\end{equation}

\begin{figure*}[htb]
\centering \includegraphics[width=12cm]{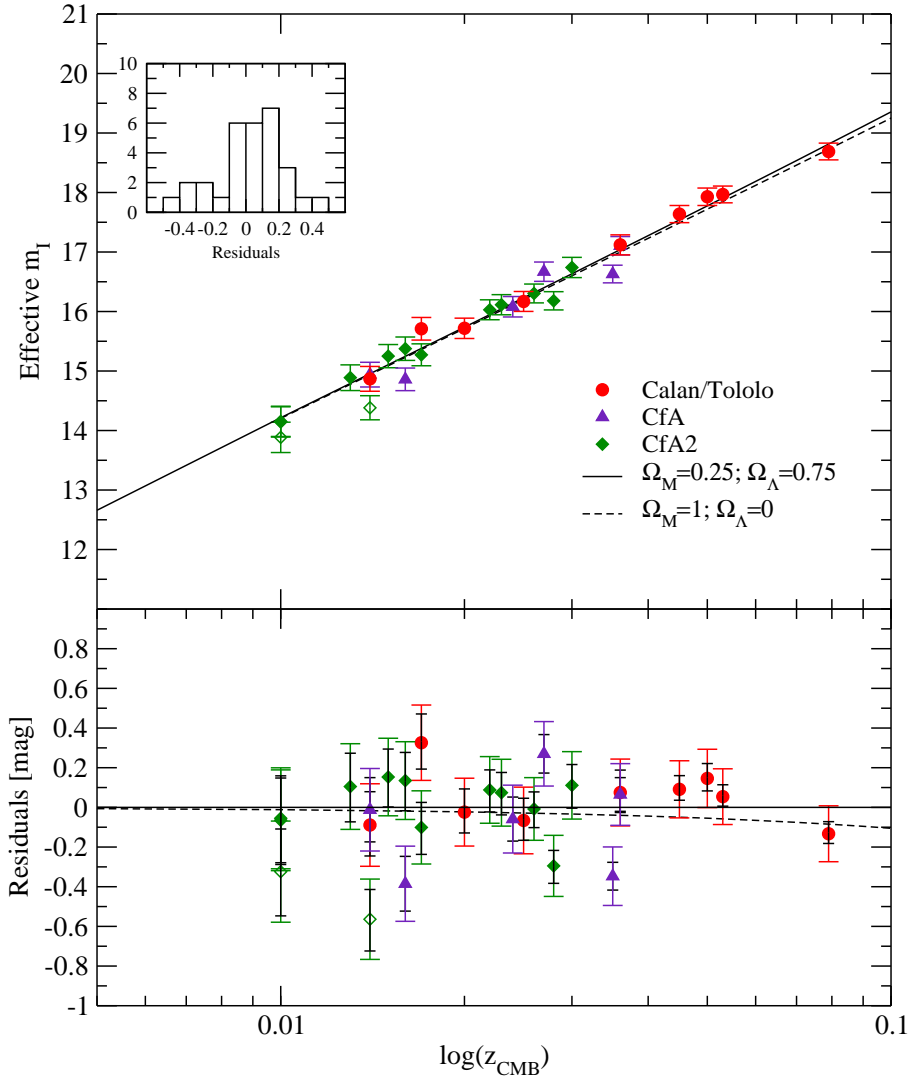}
\caption{Effective $I$-band maximum vs redshift for the nearby
  supernovae of the Calan/Tololo, CfA and CfA2 sample. The data have
  been corrected for the stretch-luminosity relation and for Milky Way
  and host galaxy extinction. The r.m.s. along the concordance model
  line is $\sigma=0.17 \pm 0.04$ mag. Subtracting the contribution of
  the average uncertainty, results in 0.13 mag estimated intrinsic
  dispersion (see text for details). SN~1998es and SN~1999dq were
  excluded from the fit (see text) and are plotted with open
  diamonds. The inset plots the histogram of the residuals. }
\label{Ihubblelow}
\end{figure*}

\noindent The effective magnitude, $m_I^{\rm eff}$ of the nearby
supernovae, listed in Table~\ref{Iband}, have been used for building
the Hubble diagram in $I$-band, shown in Fig.~\ref{Ihubblelow}. 
The inner error bars include an uncertainty in the redshifts due to
peculiar velocities of the host galaxies, assumed to be 300 km s$^{-1}$.

The solid line represents the best fit to the data for the concordance
model with fixed $\Omega_M=0.25$ and $\Omega_\Lambda=0.75$. The single
fitted parameter, $\mathcal M_I$, is defined (as in
\citet{perlmutter97}) to be

\begin{equation}
{\mathcal M}_I\equiv M_I-5\log H_0+25
\end{equation}

\noindent where $M_I$ is the $I$-band absolute magnitude for a
$B$-band stretch $s_B=1$ supernova. The value fitted is $\mathcal M_I
=-3.19 \pm 0.03$. The two redder supernovae, SN~1998es and SN~1999dq,
were excluded from the fit, and are plotted with different symbols in
Fig.~\ref{Ihubblelow}. 

In order to disentangle the intrinsic dispersion from the statistical
scatter due to the measurement uncertainties, we simulated data sets
with a dispersion given by the measurement uncertainty only. Since the
uncertainty due to peculiar motion of the host galaxy is dominant at
very low redshift ( $\sim$0.2 mag for z =0.01), we limited this
calculation to only 15 SNe with z $>$ 0.025, which correspond to a
peculiar velocity uncertainty of the same order as the measurement
uncertainties in our sample. The average of the r.m.s. measured on
each of the simulated data sets is geometrically subtracted from the
dispersion measured as r.m.s. on the data (0.17 mag), resulting in
$\sigma=0.13$ mag. We consider this an estimate of the intrinsic
dispersion of the stretch corrected $I$-band light curve maximum,
which agrees with the estimate given by \citet{Hamuyhd} using 26 SNe
of the Calan/Tololo sample. The estimated intrinsic uncertainty of
0.13 mag has been added in quadrature to the outer error bars of the
plotted data.  Note that if no correction $\alpha_I (s_B-1)$ is
applied the dispersion in the Hubble diagram becomes 0.24 $\pm$ 0.04
mag, somewhat smaller than the corresponding dispersion measured in
the ``uncorrected'' $B$-band Hubble diagram.  Moreover, we computed
the dispersion in the Hubble diagram for the three data sets
separately, and no statistically significant differences were
found.

\section{High redshift supernovae}
\label{sec:hiz}

Next, we explore the possibility of extending the Hubble diagram to
higher redshifts, where the effects of the energy density components
of the universe are, in principle, measurable. The restframe $I$-band
data available to date for this purpose are unfortunately very
limited. They consist of only three supernovae (SN~1999Q, SN~1999ff and
SN~2000fr) at redshift $z \sim 0.5$ observed in the near infrared
(NIR) J-band collected during three different campaigns conducted
using different facilities and by two different teams.  Keeping all of
these possible sources of systematic errors in mind, we include two of these
supernovae in the $I$-band Hubble diagram.

\subsection{SN~2000fr}

SN~2000fr was discovered by the Supernova Cosmology Project (SCP)
during a search for Type~Ia supernovae at redshift $z \sim 1$
conducted in the $I$-band with the CFH12k camera on the
Canada-France-Hawaii Telescope (CFHT) \citep{IAUC}. The depth of the
search allowed us to discover this $z \sim 0.5$ supernova during its
rise, about 11 rest-frame days before maximum B-band light.

The supernova  type was  confirmed with spectra  taken at the  Keck II
telescope  and  the VLT,  showing  that it  was  a  normal Type~Ia  at
$z=0.543$ (see \citet{lidman,bruch_beeth} for an extensive analysis of
the spectrum). This  supernova was followed in the  restframe $B$, $V$
and   $I$    bands   involving    both   ground   and    space   based
facilities.  Approximately one  year later,  when SN~2000fr  had faded
sufficiently,  infrared and  optical images  of the  host  galaxy were
obtained.  The  optical light  curves in \citet{rob03}  were re-fitted
using    the    improved     spectral    templates    for    computing
$k$-corrections. We found a  $B$-band stretch factor of $s_B=1.034 \pm
0.013$   and   a  time   of   $B_{\rm   max}$,   $t_{\rm  max} = \rm MJD$
51685.6. Restframe  $B-V$ measurements  at the  time of  $B_{\rm max}$
indicate that SN~2000fr  did not suffer from reddening  due to dust in
the host galaxy (see Section~\ref{sec:dust} for a more extensive
discussion).  The adopted  Milky Way  reddening is  $E(B-V)=0.030$ mag
\citep{Schlegel}.

\begin{figure}[htb]
\centering
\includegraphics[width=0.45\textwidth]{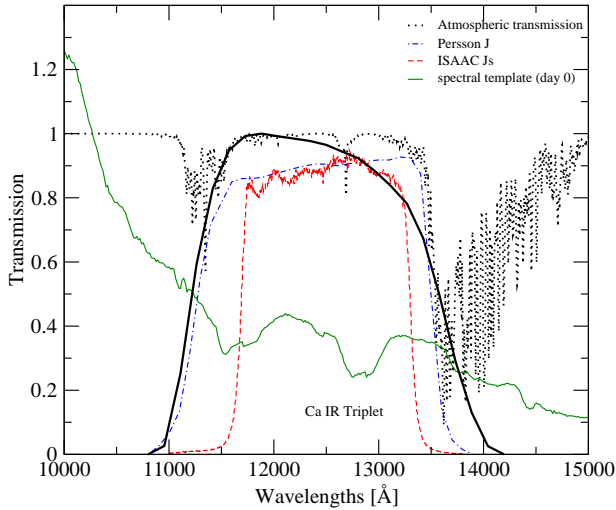}
\caption[Comparison of filters]{Comparison between the Persson $J$
filter (dashed-dotted line), the $J_s$ filter at ISAAC used
for the observations of SN~2000fr (dashed line) and the $I$-band
red-shifted to z=0.543 (heavy solid line). The atmospheric transmission
is also plotted (dotted line). The spectral template at day 0 is on an
arbitrary flux scale for readability purpose (solid line).}
\label{filters}
\end{figure}

The near-infrared data were collected with ISAAC at the VLT.  They
consist of $J_s$-band observations during three epochs and a final
image of the host galaxy without the SN (see Table \ref{tableIR}).
Each data point is composed of a series of 20 to 60 images with random
offsets between exposures. Figure \ref{filters} shows a comparison
between the Persson $J$ filter and the narrower ISAAC $J_s$ filter used for
the observations, together with the atmospheric transmission, and the
spectral template at maximum.

The advantage in using the narrower $J_s$ filter is that the
transmission of the filter is not determined by the region of strong
atmospheric absorption between 13500 and 15000 \AA. Consequently, the
zero-point is significantly more stable than that of standard
$J$. This was very useful, because all the ISAAC data were taken in
queue mode, where typically only one or two standard stars, chosen
from the list of \citet{Persson}, are observed during a night. All
data, except the reference images, were taken during photometric
nights and the difference in the zero-points from one night to the
next was less than 0.01 magnitudes.

\begin{table}[htb]
\begin{center}
\begin{tabular}{cccc}\hline
 MJD   & Epoch   &      $J_s$(mag) &  $I$(mag) \\
\hline\hline
   51685.06  & -0.33  & 22.50 $\pm$  0.09  & 23.52 $\pm$  0.10\\
   51709.02  & 15.20  & 23.57 $\pm$  0.22  & 24.52 $\pm$  0.23\\
   51731.96  & 30.07  & 23.14 $\pm$  0.15  & 23.99 $\pm$  0.16\\
\hline
\end{tabular}
\caption{Summary of $J_s$-band data for SN~2000fr. The quoted errors
are due to statistical Poisson noise and the uncertainty in the ZP
(contributing 0.01 mag). Epochs are in restframe days relative to the
date of $B$-band maximum. The restframe $I$-band is obtained through
cross-filter $k$-correction from the observed $J_s$-band to Bessel
$I$-band. The uncertainties also include the contribution from
$k$-corrections, estimated to be 0.05 mag at all epochs considered.}
\label{tableIR}
\end{center}
\end{table}

The data were reduced using both internally developed routines and the
XDIMSUM package in IRAF\footnote{IRAF is distributed by the National
Optical Astronomy Observatories, which are operated by the Association
of Universities for Research in Astronomy, Inc., under cooperative
agreement with the National Science Foundation.}. The differences between 
the two analyses are within the quoted
uncertainties. The supernova images were aligned with the host galaxy
images and the flux scaled to the one with best seeing, using the
field stars before performing PSF photometry \citep{seb_thesis}. The results are
presented in Table~\ref{tableIR}. The stated uncertainties include the
statistical Poisson noise and the uncertainty on the estimate of the
zero point, added in quadrature.

The $J_s$-band magnitude takes into account a colour term which arises
from the difference between the $J$ filter of the standard star system
and the $J_s$ filter used in ISAAC. This correction was small, $\sim$
0.012 mag.

The cross-filter $k$-correction, $K_{IJ_s}$, to convert from
$J_s$-band to rest-frame $I$-band, has been calculated following
\cite{KimGoobarPerlmutter96} using the spectral templates improved for
this work. The $k$-correction includes a term to account for the appropriate
transformation between IR and optical photometric systems, equal to
$(I-J)=0.03 (\pm 0.02)$, determined by using the Vega magnitudes in
$I$ and $J$ \citep{Bessell,Cohen}. We conservatively assume 0.05 mag
total uncertainty in the $k$-corrections (see
Section~\ref{sec:fitting}).

\subsection{SN~1999ff}

SN~1999ff was discovered by the High-Z Supernova Search Team (HZSST)
during a search conducted at CFHT using the CFH12k camera in the $I$-band
\citep{tonry}.\footnote{Another supernova, SN~1999fn, was followed in
$J$-band by the HZSST during the same search. However since it was found
in a highly extincted Galactic field, E(B-V)=0.32 mag, and since it
was strongly contaminated by the host galaxy, we did not include it in
our analysis.} The supernova was confirmed spectroscopically as a
Type~Ia at redshift $z=0.455$. The adopted Milky Way reddening is
$E(B-V)=0.025$ mag \citep{Schlegel}.

$J$-band observations, corresponding to restframe $I$-band, reported
in \citet{tonry}, were taken at Keck using NIRC at two epochs
only. The $J$-band filter that was used for these observations is very
similar to the ISAAC $J_s$, shown in Fig.~\ref{filters}. We have
used the published photometry, and, for consistency with the
treatment of both the low redshift supernovae and SN~2000fr, we
computed the $k$-corrections using the improved spectral templates. We
found differences with the results published in \citet{tonry}, due to
the use of an incorrect filter in the originally published results.
(However, the $k$-corrections calculated as part of the MLCS distance
fits to this object were done with the correct filter (Brian Schmidt,
private communication)).  The $I$-band magnitudes were also corrected
for the offset found between the optical and IR systems, as explained
in the previous section. The restframe $I$-band magnitudes obtained
this way are reported in Table~\ref{phot99ff}. The published optical
$R$-band data were used to fit the restframe $B$-band light curve
using the stretch method. Our time of maximum was within 1 day of the
\citet{tonry} value, with a best fit for the stretch $s_B=0.80 \pm
0.05$.

\begin{table}[htb]
\begin{center}
\begin{tabular}{ccc}\hline
 $\rm MJD$ & Epoch    &       I(mag)  \\
\hline\hline
51501.29 &  5.01  & 23.57 $\pm$  0.11  \\
51526.31 & 22.21  & 24.06 $\pm$  0.24  \\
\hline
\end{tabular}
\caption{Summary of IR data for SN~1999ff. Epochs are in restframe
  days relative to the date of the $B$-band maximum ($t_{\rm max}={\rm
  MJD} 51494.8$); restframe $I$-band magnitudes are computed applying
  $k$-corrections to the observed $J$-band data published in
  \citet{tonry}. The uncertainties also include the contribution from
  $k$-corrections, estimated to be 0.05 mag at all epochs considered.}
\label{phot99ff}
\end{center}
\end{table}

\subsection{SN~1999Q}
\label{sec:99q}

SN~1999Q was discovered by the HZSST using the CTIO 4m Blanco
Telescope and was spectrally confirmed to be a Type~Ia SN at $z=0.46$
\citep{IAUC 7097}. The adopted Milky Way reddening is $E(B-V)=0.021$
magnitudes \citep{Schlegel}.

SN~1999Q was observed in the $J$-band over five epochs, the first with
SofI on the ESO NTT and the following four with NIRC at the Keck Telescope
\citep{riess99q}.  We recomputed the $k$-corrections using our new
spectral template (as we did for SNe 2000fr and 1999ff) and we find a
difference of up to 0.15 magnitudes between our $k$-corrections and
those published in \citet{riess99q}.

A fit to the published restframe $I$-band data of SN~1999Q shows that
it is a 4 standard deviation outlier in the $I$-band Hubble diagram.
In order to investigate its faintness, we re-analysed the publicly
available SofI data and found $J$=22.63 $\pm$ 0.15 mag, which is
significantly brighter than the published value, $23.00 \pm 0.14$ mag
\citep{riess99q}. Due to this large discrepancy, we decided to not
include this SN in the rest of the analysis.

\subsection{Light Curve fits for the high redshift supernovae.}
\label{hizfit}

The $I$-band light curves of the high redshift supernovae are not as
well sampled in time as the low redshift sample analysed in Section
\ref{sec:ibandfit}.  There are only few data points for each SN,
making it impossible to perform the 5 parameter fit. Thus, we used the
results of the fit of the local sample of supernovae to build a set of
42 $I$-band templates, which in turn have been used to fit the high
redshift SN light curves.

\begin{figure}[htb]
\centering \includegraphics[width=8cm]{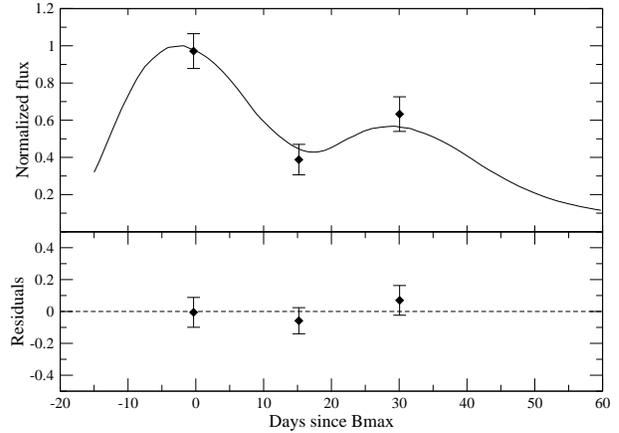}
\caption{$I$-band fit for SN~2000fr. Out of 42 $I$-band templates, the
  best fit was obtained with the template of SN~1992bc. The fit was
  performed with only one free parameter, the peak magnitude, $I_{\rm
  max}=23.48 \pm 0.08$ mag. Supplemental data from the $B$-band (not
  shown) is used to fix the date of $B$-maximum.}
\label{sn2000fr_I}
\end{figure}

\begin{figure}[htb]
\centering \includegraphics[width=8cm]{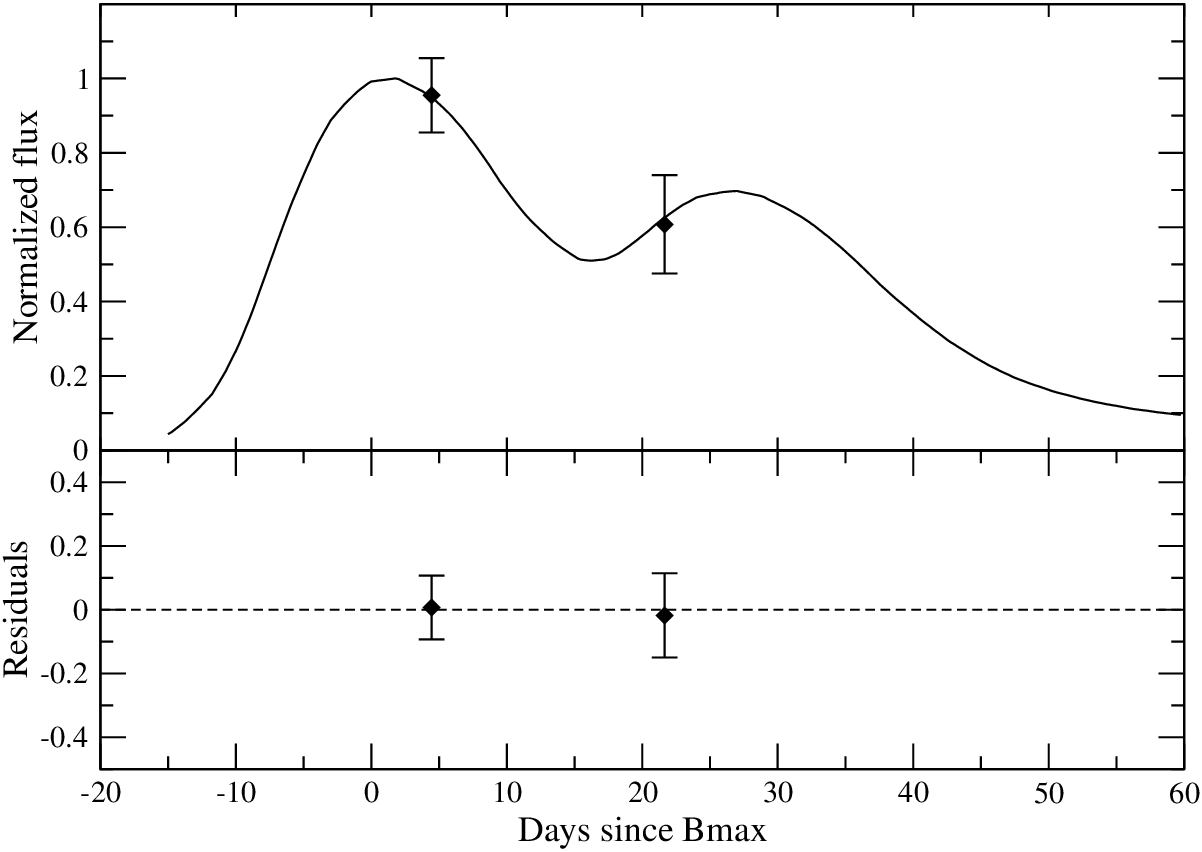}
\caption{$I$-band fit for SN~1999ff.  Out of 42 $I$-band templates,
  the best fit was obtained with the template of SN~1996bl. The fit
  was performed with only one free parameter, the peak magnitude,
  $I_{\rm max}=23.55 \pm 0.10$ mag. Supplemental data from the
  $B$-band (not shown) is used to fix the date of $B$-maximum.}
\label{sn1999ff_I}
\end{figure}

The best fit light curve for each of the 42 supernovae in our
low-redshift sample can be viewed as defining an $I$-band template.
The high redshift supernovae are fit to each template with a single
free parameter, $I_{\rm max}$, the absolute normalisation of the
template.  The time of $B_{\rm max}$ is obtained from the literature
(SN~1999ff) or from our own $B$-band light curve fits (SN~2000fr).
The best-fitting low-redshift $I$-band template fixes the date of the
I-band maximum relative to the date of the $B$-band maximum. A
$\chisq$ comparison was used to choose the best low redshift
template. 
Figs.~\ref{sn2000fr_I} and \ref{sn1999ff_I} show the comparison of the
data with the best fit template for each of the supernovae.
Table~\ref{high-z} gives the results of the fit together with
redshift, the number of data points, the template giving the best fit
and the $\chisq$. As there are only a few data points for each SN, the
$\chi^2$ parameter has little significance for estimating the goodness
of the fits. Thus, to estimate the possible systematic error in the
measured peak magnitude from the selection of the light curve
template, we computed the r.m.s. of the fitted $I_{max}$ of all the
light curve templates satisfying $\chi^2 \leq \chi_{min}^2+3$. This
possible systematic uncertainty is reported also in Table~\ref{high-z}. For
both SNe this is quite small, and compatible with the scatter due to
the statistical uncertainties, thus, it is a conservative estimate.

\begin{table*}[htb]
\begin{center}
\begin{tabular}{ccclclcc}\hline
 SN   & $z$ & $s_B$ & $n$ & $I_{\rm max}$ & template & $\chisq$ & $A_J^{MW}$\\
\hline\hline
SN~2000fr & 0.543 & 1.034 $\pm$ 0.011 & 3 & 23.48 $\pm$ 0.08 $\pm$ 0.04 & SN~1992bc & 1.04 & 0.027\\
SN~1999ff & 0.455 & 0.80 $\pm$ 0.05   & 2 & 23.55 $\pm$ 0.10 $\pm$ 0.08 & SN~1996bl & 0.05 & 0.022\\
\hline
\end{tabular}
\caption{List of the high redshift Type~Ia SNe used in this work.
  Columns are: IAU name, redshift, number of data points used in the fit,
  magnitude of the peak resulted from the fit (both statistical and
  systematic uncertainties are given) before Milky Way extinction
  correction, best fit template, $\chisq$ of the fit, Milky Way
  extinction in the $J$-band.}
\label{high-z}
\end{center}
\end{table*}

\subsection{Monte-Carlo test of the fitting method}

A Monte-Carlo simulation was run in order to test the robustness of
the fitting method applied to the high redshift SNe. The measurement
uncertainties were used to generate a set of 1000 SNe, with data
points randomly distributed around the real data and at the same
epochs as the data. All the simulated data sets were in turn fitted
with the 42 templates and the one giving the minimum $\chi^2$ was
selected for each simulation. The distribution of the fitted
parameters in each of the simulated data sets around the true values,
fitted on the experimental data, was studied to check for systematic
uncertainty in the fitting procedure. This was found to be robust,
always selecting the same template as the one giving the best fit for
both SNe. No bias was found, therefore confirming the peak
magnitude fitted with this method. The uncertainty in $I_{\rm max}$
reported in Table~\ref{high-z} was consistent with the dispersion in
the distribution of $I_{\rm max}$ measured from the simulations.

\section{The $I$-band Hubble diagram up to $z \sim 0.5$}

The $I$-band peak magnitudes of the high redshift supernovae reported
in Table~\ref{high-z} were corrected for Milky Way extinction. Note
that both SN~1999ff and SN~2000fr have been reported not to suffer
from extinction from their host galaxies.

\begin{figure*}[htb]
\centering \includegraphics[width=12cm]{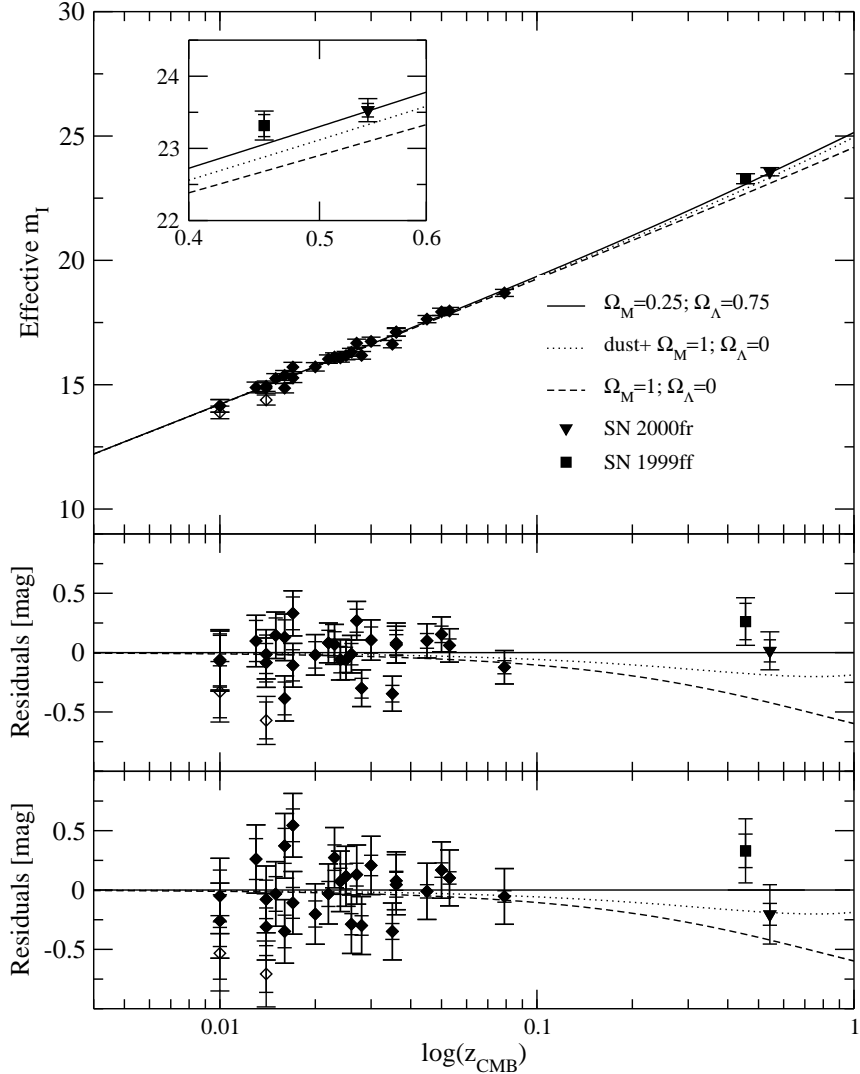}
\caption{Effective $I$-band maximum versus redshift for the nearby
  supernovae of the Calan/Tololo, CfA and CfA2 sample, together with
  two supernovae at redshift $\sim 0.5$ for case $c$ (top panel),
  residuals to the ($\Omega_M,\Omega_\Lambda$)=(0.25,0.75) model for
  case $c$ (middle panel) and case $a$ (bottom panel). SN~1998es and
  SN~1999dq were excluded from the fit (see text) and are plotted with
  open diamonds. Only the low-redshift SNe have been extinction
  corrected.  The host galaxy extinction correction would make the two
  high-redshift data points closer to each other but with larger error
  bars.}
\label{hubbleI}
\end{figure*}

The Hubble diagram has been built both with and without
width-luminosity correction (case $c$ and case $a$ respectively),
where the systematic uncertainties on the peak magnitudes of the
distant supernovae, listed in Table~\ref{high-z}, are added in
quadrature to the statistical uncertainties. Cases $b$ and $d$ are
like $a$ and $c$ but neglect the systematic uncertainties from
Table~\ref{high-z}.  Figure~\ref{hubbleI} shows the extended Hubble
diagram (case $c$), where an intrinsic uncertainty of 0.13 mag has
been added in quadrature to the measurement errors of the plotted
data. The solid line represent the best fit to the nearby data for the
concordance model $\Omega_M=0.25$ and $\Omega_{\Lambda}=0.75$. Also
plotted is the model for $\Omega_M=1$ and $\Omega_{\Lambda}=0$ (dashed
line), and a flat, $\Lambda=0$ universe in the presence of a
homogeneous population of large dust grains in the intergalactic (IG)
medium able to explain the observed dimming of Type~Ia SNe at $z \sim
0.5$ in the $B$-band (dotted line)
\citep{aguirre1999a,aguirre1999b}. The bottom panel shows the
residuals obtained for case $a$.  Table~\ref{HDchi2} lists the
$\chisq$ values for the high redshift SNe for each of the models.  The
$\Omega_\Lambda$-dominated cosmology is formally favored over the
other two models at the $>$2$\sigma$ level.  However, two high
redshift supernovae obviously do not provide the full gaussian
distribution that would confirm this result.


\begin{table}[htb]
\begin{center}
\begin{tabular}{lccccc}\hline
$(\Omega_M,\Omega_\Lambda)$ & $\chisq _a$ & $\chisq _b$ & $\chisq _c$ & $\chisq _d$ \\
\hline\hline
(0.25,0.75)    & 2.18 & 2.36  & 1.16  & 1.44 \\
(1,0)          & 7.56 & 8.44  & 15.56  & 18.30 \\
$(1,0)_{dust}$ & 3.52 & 3.96  & 5.36 & 6.48 \\
\hline
\end{tabular}
\caption{$\chisq$ (for 2 {\it dof}) of each model to the high
  redshift data, without stretch correction and with systematic
  uncertainties added in quadrature ($\chisq _a$), neglecting the
  systematic uncertainties ($\chisq _b$), with stretch correction and
  adding the systematic uncertainties in quadrature ($\chisq_c$) or
  neglecting them ($\chisq_d$).}
\label{HDchi2}
\end{center}
\end{table}

Systematic uncertainties in the method used here also cannot be
extensively explored with only two supernovae.
Some uncertainties are specific to the sample considered
here. The different fitting methods applied to the restframe $I$-band
light curve for the low and high redshift samples can be easily
overcome if distant supernovae are followed at NIR wavelengths with
better time coverage. Both the low and high redshift samples used in
this analysis are rather heterogeneous, as they were collected from
different data sets. Future data sets collected with a single
instrument would naturally solve this problem.

\section{SN~Ia colours and intergalactic dust}
\label{sec:dust}

Multi-colour photometry allows one to search for non-standard
dust having only a weak wavelength dependence, such as a homogeneous
population of large grain dust, as proposed by
\citet{aguirre1999a,aguirre1999b}. 

If we assume that grey dust is responsible for the dimming of SNe~Ia
in the $B$-band at $z\sim0.5$, we can calculate the expected extinction in
other filters and compute the resulting colours. Following
\citet{ariel}, we use the SNOC Monte-Carlo package \citep{snoc}
for two cases of the total to selective extinction ratio - $R_V= 4.5$
and $9.5$. We assume that the dust is evenly distributed between
us and the SNe in question and we assume a flat cosmological model
with a zero cosmological constant.

\begin{figure}[bht]
\centering
\includegraphics[width=8cm]{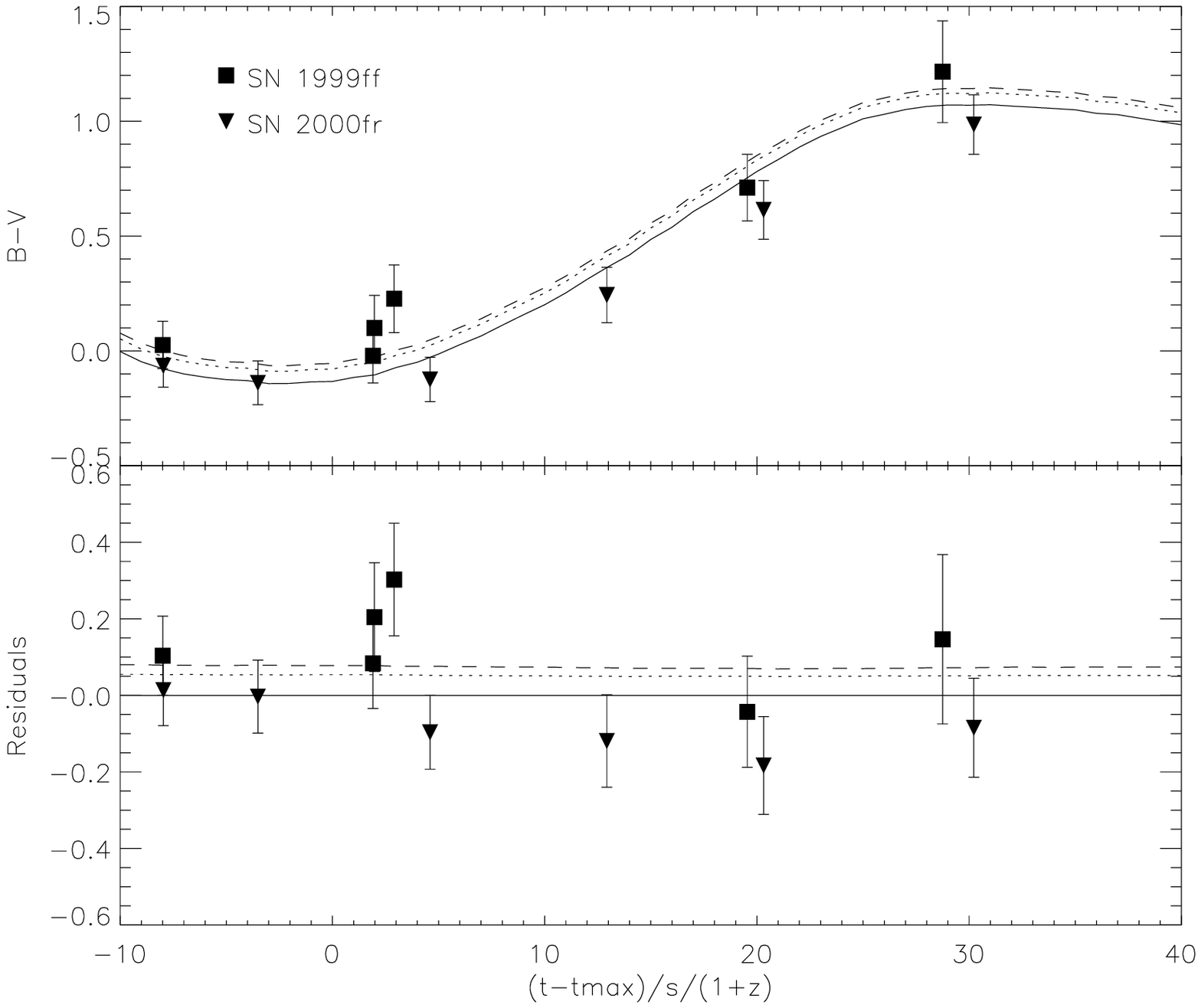}\\
\includegraphics[width=8cm]{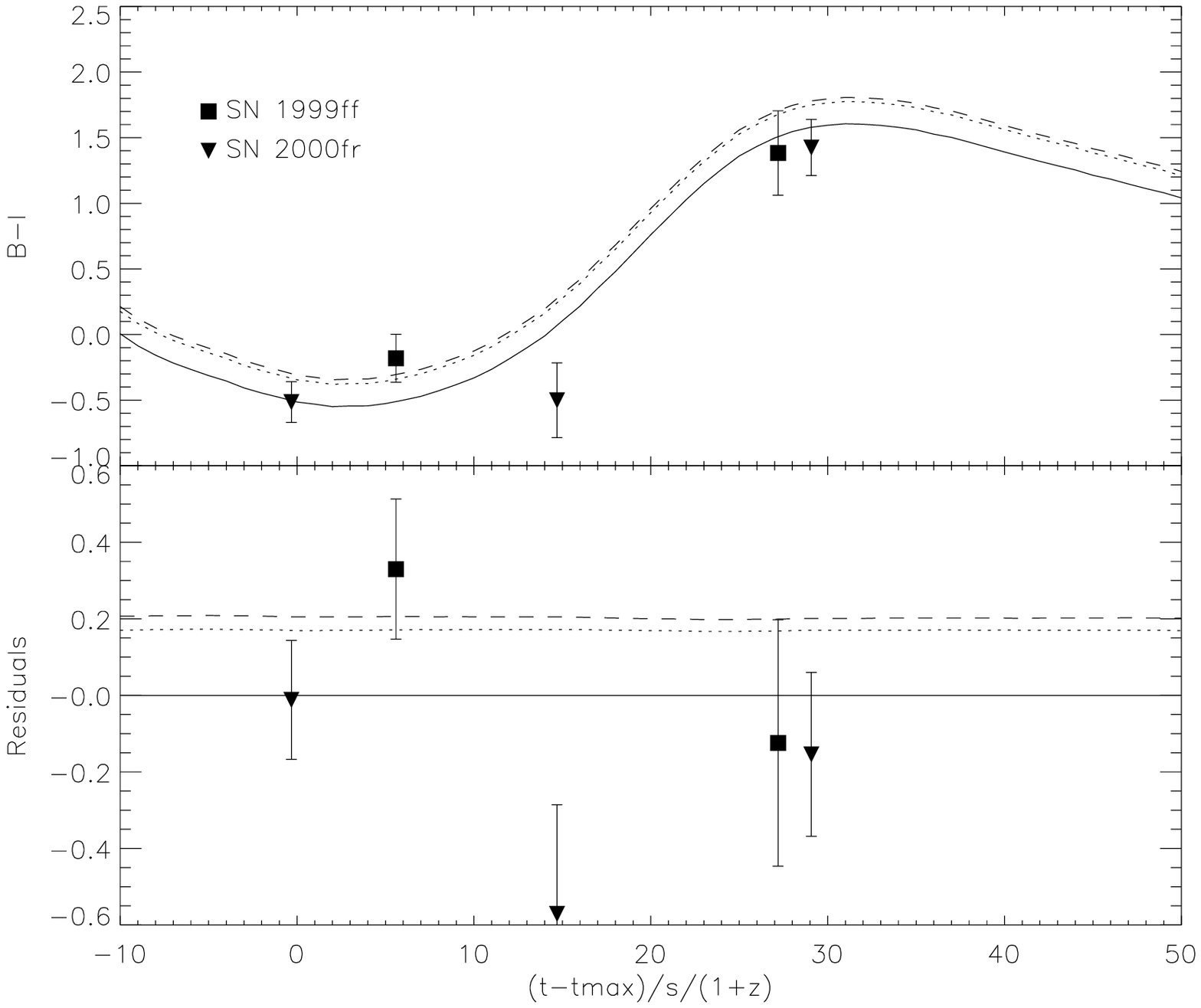}\\
\caption{The evolution in the colour of SN~1999ff (squares) and
SN~2000fr (triangles), $B-V$ (top panel) and $B-I$ (bottom panel),
compared to the colour evolution of the average SNe~Ia in a $\Lambda$
dominated universe (solid line) and a $\Omega_M =1$, $\Omega_\Lambda
=0$ universe with presence of grey dust with $R_V$=4.5 (dashed line)
and $R_V$=9.5 (dotted line) for $z=0.5$.} 
\label{colors}
\end{figure}

The measured $B-I$ and $B-V$ colours of SN~1999ff and SN~2000fr,
corrected only for Milky-Way extinction, are presented in
Tables~\ref{tab:BIcolours} and \ref{tab:BVcolours} and plotted in
figure~\ref{colors}.
The error bars include the contribution of the intrinsic colour
dispersion. The expected evolution in the $B-I$ and $B-V$ colours of an average
SNe~Ia in the concordance model and in the two models with grey dust
and without a cosmological constant at $z=0.5$ are also shown.

The $\chisq$ has been computed for both $B-V$ and $B-I$ evolution for
SN~1999ff and SN~2000fr, and for both supernovae together (see
Table~\ref{chi2table}). The correlations between SN colours at
different epochs found in \citep{nobili2003} were taken into account.
However, we note that, although this correlation should be taken into
account in the calculations, neglecting it would not change
significantly the conclusions of the analysis.  Although individual
supernovae give $\chisq$ values that would seem to distinguish between
the models, the combined results disfavour such
conclusions.

\begin{table}[htb]
\begin{center}
\begin{tabular}{cc}\hline
Epoch &   $B-I$  \\
\hline\hline
SN~2000fr  & \\
\hline
-0.32  &  -0.51 $\pm$   0.12 \\
14.70  &  -0.50 $\pm$   0.24 \\
29.08  &   1.42 $\pm$   0.17 \\
\hline
SN~1999ff  & \\
\hline
 5.59  &  -0.18 $\pm$   0.13 \\
27.20  &   1.38 $\pm$   0.24 \\
\hline
\end{tabular}
\caption{Restframe $B-I$ colours in magnitudes for the two high
  redshift SNe. The Epoch is in restframe days relative to the
  $B$-band maximum, divided by the $B$-band stretch.}
\label{tab:BIcolours}
\end{center}
\end{table}

\begin{table}[htb]
\begin{center}
\begin{tabular}{cc}\hline
Epoch &  $B-V$ \\
\hline\hline
SN~2000fr & \\
\hline
  -7.97  &  -0.06 $\pm$   0.05\tablenotemark{a}  \\
  -3.51  &  -0.14 $\pm$   0.05\tablenotemark{a}  \\
   4.60  &  -0.12 $\pm$   0.05  \\
  12.93  &   0.24 $\pm$   0.08  \\
  20.31  &   0.61 $\pm$   0.07  \\
  30.22  &   0.99 $\pm$   0.09  \\
\hline
SN~1999ff & \\
\hline
 -7.99  &   0.03 $\pm$   0.08\tablenotemark{a}  \\
   1.91  &  -0.02 $\pm$   0.09  \\
   1.98  &   0.10 $\pm$   0.12  \\
   2.91  &   0.23 $\pm$   0.12  \\
  19.55  &   0.71 $\pm$   0.09  \\
  28.75  &   1.22 $\pm$   0.20  \\
\hline
\end{tabular}
\end{center}
\caption{Restframe $B-V$ colours in magnitudes for SN~1999ff and
SN~2000fr.  The Epoch is in rest frame days relative to the $B$-band
maximum, divided by the $B$-band stretch.}  \tablenotetext{a}{The data
are not included in the analysis because they are out of the range in
which \citet{nobili2003} studied colour
correlations.}
\label{tab:BVcolours}
\end{table}

\begin{figure*}[hbt]
\centering
\includegraphics[width=8cm]{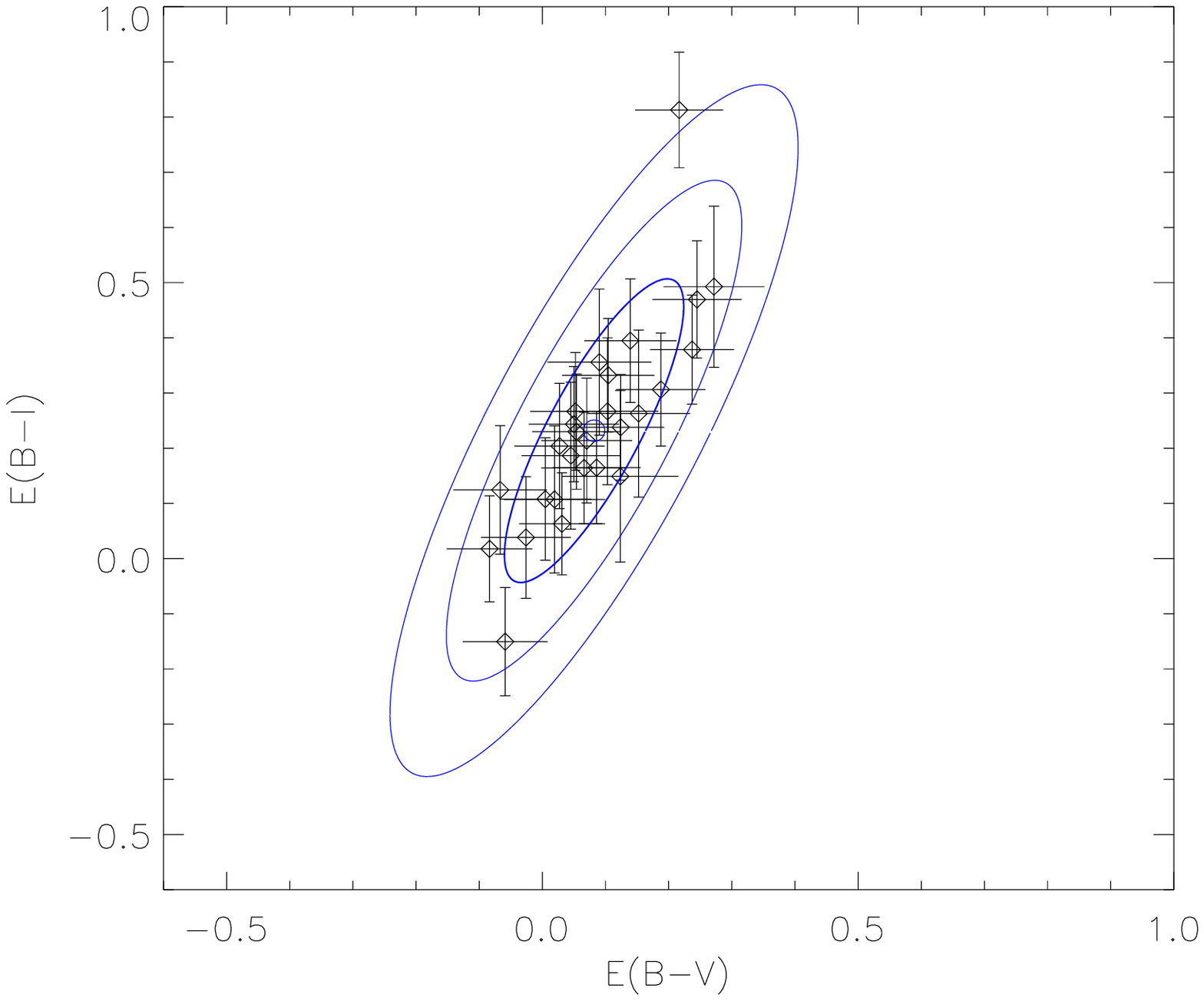} 
\includegraphics[width=8cm]{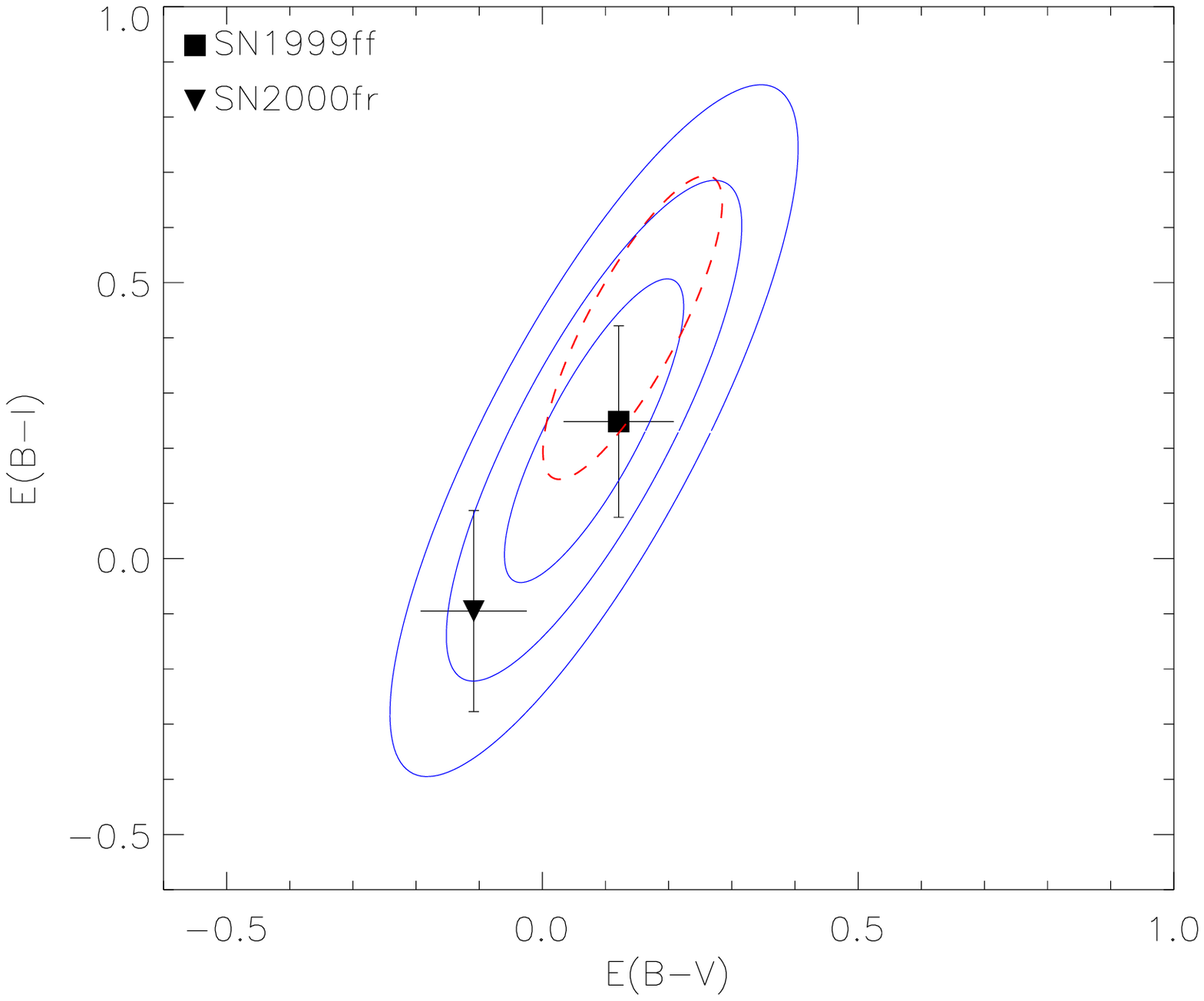} 
\includegraphics[width=8cm]{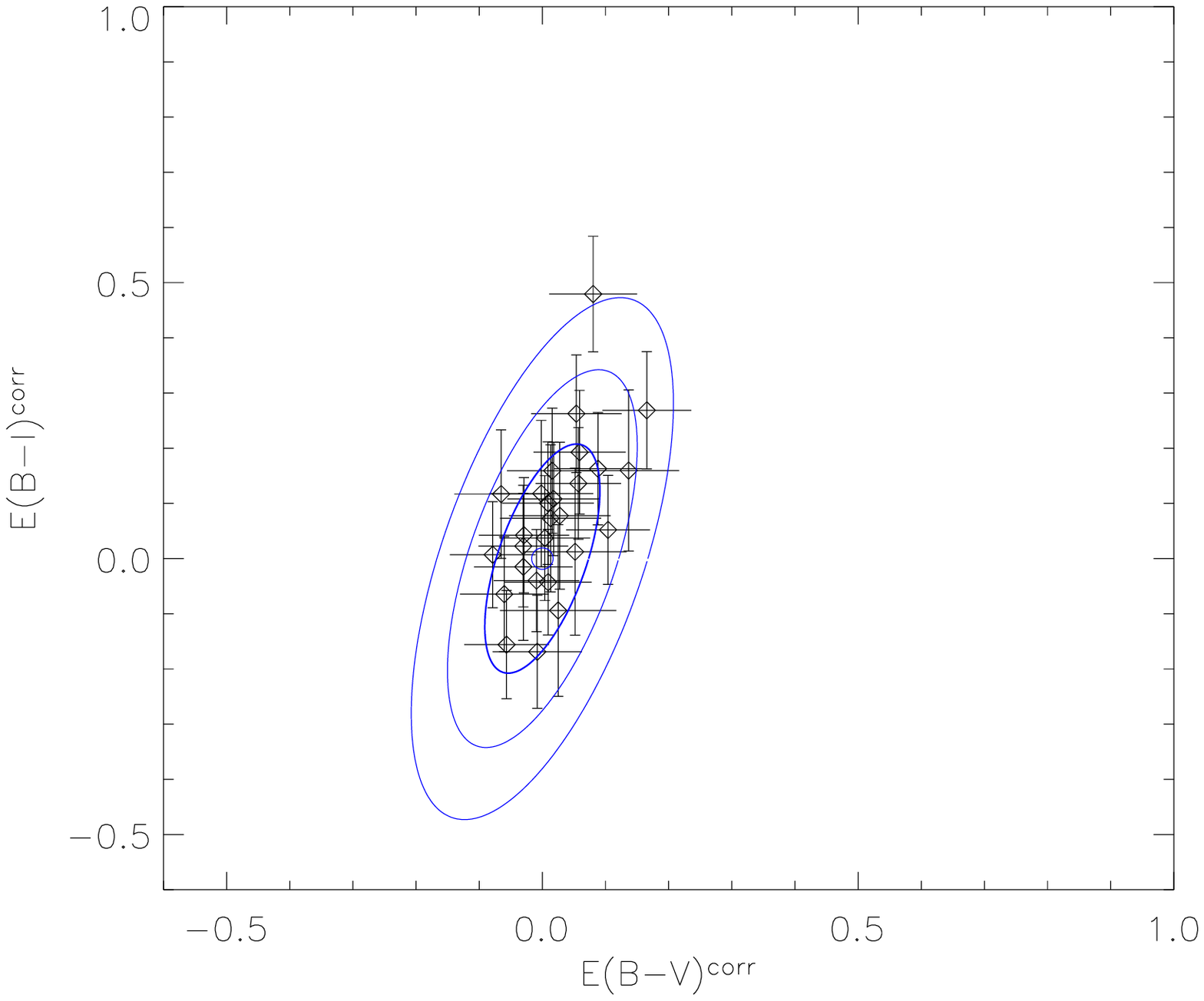} 
\includegraphics[width=8cm]{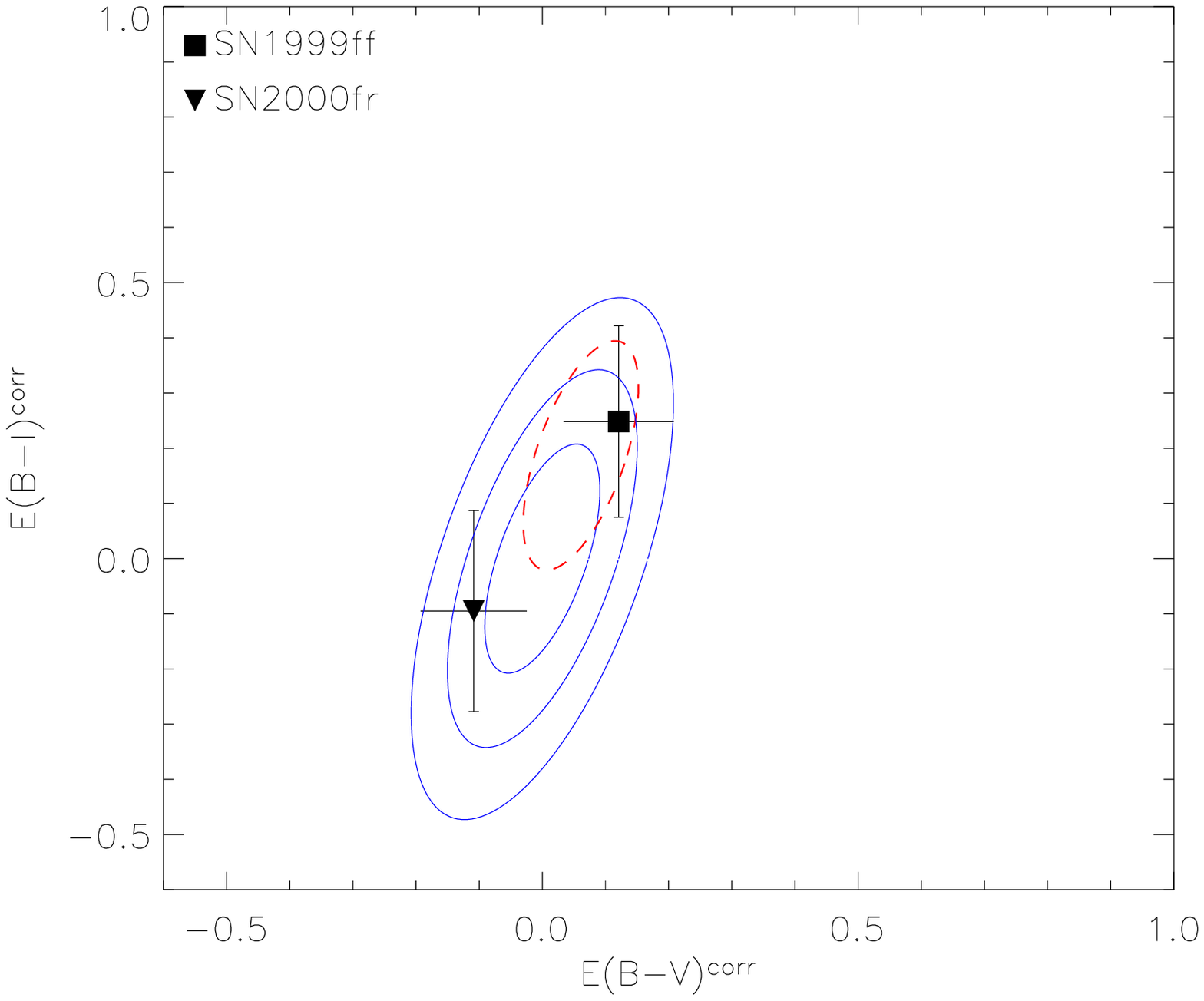} 
\caption{Left-hand panels: distribution of combined colour
  measurements of nearby SNe~Ia in the $E(B-I)$ vs $E(B-V)$ plane, not
  corrected (top panel) and corrected (bottom panel) for the host
  galaxy extinction. The solid contours incorporate 68.3\%, 95.5\% and
  99.7\% of the sample. Right-hand panels: SN~1999ff and SN~2000fr
  compared to the distribution defined by nearby SNe in the case of no
  IG dust (solid ellipses), and in the case of IG dust with $R_V=9.5$
  (dashed ellipse).  For clarity only the 68.3\% level has been
  plotted for the dust distribution. }
\label{ellipses}
\end{figure*}

To make our test for grey dust more effective, a different approach
was followed. The method of least squares has been used to combine
colour measurements along time for each supernova (see Cowan, 1998,
p.106 for details). The residuals between the data and the models are
averaged with a weight that is determined from the covariance matrix.
In the following, we refer to $E(X-Y)$ to describe the colour excess
of any supernova with respect to the average $X-Y$ colour of nearby
SNe~Ia, as derived in \citep{nobili2003}. First we applied this
method to all local supernovae and used the results to establish the
expected distribution in the $E(B-I)$ vs $E(B-V)$ plane, as showed in
Fig.~\ref{ellipses}.

As the high redshift SNe were not corrected for host galaxy
extinction, we computed the colour distribution of nearby SNe~Ia for
two cases: the left-hand panels represent the distribution of colour
excess of 27 nearby SNe not corrected (top panel) and corrected
(bottom panel) for host galaxy extinction. Spectroscopically peculiar
SNe have been excluded from the analysis. The projection of the
ellipses on each colour axis is the estimated standard deviation in
that colour and the inclination is defined by the linear Pearson
correlation coefficient computed on the same data sample. The solid
contours represent 68.3\%, 95.5\% and 99.7\% probability.

The right-hand panels in Fig.~\ref{ellipses} show the combined values
of colour excess, $[E(B-V),E(B-I)]$, for the high redshift supernovae:
$[0.12 \pm 0.09, 0.25 \pm 0.17]$ for SN~1999ff and $[-0.11 \pm 0.08,
-0.10 \pm 0.18]$ for SN~2000fr.  These are compared to the local
supernova distribution (solid lines), that represent the distribution
expected in the absence of IG dust. Also plotted is the 68.3\% level
of the expected distribution in presence of grey dust with $R_V= 9.5$,
represented by the ellipse (dashed line) that is displaced by
(0.06,0.19) from the no-dust model. Only the case of $R_V= 9.5$ has
been plotted for readability reasons, given the small difference
between the two dust models. Note that this is the closer to the
no-dust model.  
The ellipse corresponding to $R_V= 4.5$ would be displaced by
(0.03,0.04), respectively in $E(B-V)$ and $E(B-I)$, from the $R_V=
9.5$ model.

We computed the $\chisq$ of the high redshift data for all three
models, for two cases: in the first case, the nearby SNe~Ia are
corrected for extinction by dust in the host galaxy, and, in the
second case, they are not (bottom and top panels of
Fig.~\ref{ellipses}). For each model, we sum the $\chisq$ contribution
from all SNe, taking into account the correlation found between
$E(B-V)$ and $E(B-I)$ in the nearby sample.  In the first case, the
reduced $\chisq$ (for 4 degrees of freedom) are 0.63, 1.20, and 0.97
for the no-dust, IG dust with $R_V=4.5$ and IG dust with $R_V=9.5$
models respectively. In the second case, the reduced $\chisq$ are
0.62, 1.55 and 1.32 respectively. We note that both the intrinsic
dispersion in the colours of the nearby data, and the uncertainties in
the colours of the high redshift SNe, have been taken into account in
computing the $\chisq$.  The statistical significance of these results
is very limited, and should only be taken as an example of the method
developed here.  Moreover, the possibility for this analysis to be
affected by systematic effects is not negligible. Increasing the
sample and the time sampling for each object would allow us not only
to improve the significance of our statistic, but it will also be a
means to identify and quantify systematic effects involved.

A Monte Carlo simulation was used to estimate the minimum sample size
needed to test for the presence of homogeneously distributed grey dust
in the IGM. SNe colours were generated following the binormal
distribution defined by the nearby SNe~Ia sample. Under the assumption
that the systematic effects are negligible and an average measurement
uncertainty of 0.05 mag in both $E(B-V)$ and $E(B-I)$, we found that a
sample of at least 20 SNe would be needed to exclude the IG dust model
with $R_V=9.5$ at the 95\% C.L. Note that the average measurement
uncertainty of 0.05 mag can be achieved with different strategies.
Currently, the uncertainties on the individual measurements give the
main contribution to the colour uncertainties.  A good sampling would
allow us to better identify and quantify currently unidentified
systematic effects which may possibly be affecting the current
analysis.

\begin{table}\begin{center}
\begin{tabular}{lll}
\hline
      &  $\chi^2_{B-V}$/dof & $\chi^2_{B-I}$/dof \\
\hline
SN~2000fr  & & \\
\hline
no dust, $(\Omega_M,\Omega_\Lambda)$=(0.25,0.75)    & 2.33/4 & 2.08/3 \\
dust $R_v=9.5$, $(\Omega_M,\Omega_\Lambda)$=(1,0) & 4.24/4 & 3.94/3 \\
dust $R_v=4.5$, $(\Omega_M,\Omega_\Lambda)$=(1,0) & 5.29/4 & 4.50/3 \\
\hline
SN~1999ff & & \\
\hline
no dust , $(\Omega_M,\Omega_\Lambda)$=(0.25,0.75)   & 6.05/5 & 3.93/2 \\
dust $R_v=9.5$, $(\Omega_M,\Omega_\Lambda)$=(1,0) & 4.69/5 & 2.06/2 \\
dust $R_v=4.5$, $(\Omega_M,\Omega_\Lambda)$=(1,0) & 4.31/5 & 1.89/2 \\
\hline
SNe combined & & \\\hline
no dust , $(\Omega_M,\Omega_\Lambda)$=(0.25,0.75)   & 8.38/9 & 6.01/5 \\
dust $R_v=9.5$, $(\Omega_M,\Omega_\Lambda)$=(1,0) & 8.93/9 & 6.00/5 \\
dust $R_v=4.5$, $(\Omega_M,\Omega_\Lambda)$=(1,0) & 9.59/9 & 6.39/5 \\
\hline
\end{tabular}\end{center}
\caption{$\chi^2$ computed for the 3 different models and colours for
  each of the supernovae and for their combination.}
\label{chi2table}
\end{table}

\section{Test for SN brightness evolution}
\label{sec:evolution}

Evolution of the properties of the progenitors of SNe~Ia with redshift
has often been proposed as an alternative explanation for the observed
dimming of distant SNe. This is based on the assumption that older galaxies
show different composition distribution than younger ones, e.g. an
increased average metallicity, resulting in different
environmental conditions for the exploding star. A simple way to test
for evolution is to compare properties of nearby SNe with distant
ones. This will not prove that there is no evolution, but it will
exclude it on a supernova-by-supernova or property-by-property basis,
 always finding counterparts of distant events in the local sample.

In this work we compared the colours of nearby and distant supernovae
(primarily to test presence of ``grey'' dust). Although the size of
the high redshift sample is very limited, our results do not show any
evidence for evolution in the colours of SNe~Ia.  Furthermore, the
correlation found between the intensity of the secondary peak of
$I$-band light curve and the supernova luminosity give an independent
way of testing for evolution. The restframe $I$-band light curve of
the high redshift supernovae were all best fitted by templates showing
a prominent second peak, i.e. inconsistent with the intrinsically
underluminous supernovae. Note that the data presented here for
SN~2000fr show for the first time a case where the secondary peak is
unambiguously evident in the data even prior to the light curve
fit. Table~\ref{chi2sublum} lists the $\Delta\chi^2$ for the fit of
the high redshift SNe to the templates of the two underluminous
SN~1991bg and SN~1997cn, relative to the best fit. The $\chi^2$ values
are significantly larger than the best fit value.

\begin{table}\begin{center}
\begin{tabular}{llcc}
\hline
   &  $n$ &  SN~1991bg &  SN~1997cn\\
\hline
SN~2000fr  & 3 & 24.03 & 21.73\\
SN~1999ff  & 2 & 3.28 & 2.72\\
\hline
\end{tabular}\end{center}
\caption{$\Delta\chi^2$ for the fit of the high redshift SNe to the
  templates of the two underluminous SNe relative to the best fits
  (which are ``normal'' SN templates). $n$ is the number of data
  points used in the one-parameter fit (see discussion in section
  \ref{hizfit}).}
\label{chi2sublum}
\end{table}

\section{Summary and conclusions}
\label{sec:conclusion}
In this work we have investigated the feasibility and utility of using
restframe $I$-band observations for cosmological purposes.

We have developed a five parameter light curve fitting procedure which
was applied successfully to 42 nearby Type~Ia supernovae. The fitted
light curves were used to build a set of templates which include a
broad variety of shapes. We have found correlations between the fitted
parameters, in particular between the time of the secondary peak and
the $B$-band stretch, $s_B$. Moreover, a width-luminosity relation was
found between the peak $I$-band magnitude and the $B$- and $I$-band
stretches ($s_B$ and $s_I$).

We built a restframe $I$-band Hubble diagram using 26 nearby
supernovae at redshifts $0.01 \le z \le 0.1$, and measured an r.m.s.
of 0.24 mag, smaller than the uncorrected dispersion corresponding to
restframe $B$-band. The width-luminosity relation was used to reduce
the r.m.s. to 0.17 $\pm$ 0.03 mag (including measurement errors),
corresponding to an intrinsic dispersion of 0.13 mag. Differences
between the three data samples are also discussed.

$J$-band measurements of one new high redshift supernova plus
published data of another were used to extend the Hubble diagram up to
$z \sim 0.5$. The restframe $I$-band light curves of the $z \sim 0.5$
supernovae were fitted with templates that were built from the nearby
SNe~Ia, as the five parameter fit method could not be used for the
poorly sampled high redshift light curves. The peak $I$-band magnitude
of the high redshift SNe was compared to three different sets of
cosmological parameters. The ``concordance model'' of the universe,
($\Omega_M,\Omega_\Lambda$)=(0.25,0.75), is formally found in better
agreement with the data than the other models at the $>$2$\sigma$
level. However, the small sample size does not yet allow strong
conclusions to be drawn.

Alternative explanations for the observed dimming of supernova
brightness, such as the presence of grey dust in the IG medium or
evolutionary effects in the supernova properties have also been
addressed. Both the $I$-band Hubble diagram and multi colour
photometry have been used for testing grey dust. Although no firm
limits on the presence of grey dust could be set, this study shows
that with higher statistics, the restframe $I$-band measurements could
provide useful information on cosmological parameters, including tests
for systematic effects. A Monte Carlo simulation indicates that a
sample of at least 20 well observed SNe~Ia would be enough for setting
limits through the multi-colour technique used in this paper.  A
similar technique, using QSO instead of SNe~Ia, was successfully used by
\citet{2003JCAP...09..009M} to rule out grey dust as being the sole
explanation for the apparent faintness of SNe~Ia at $z \sim 0.5$.

Possible systematic uncertainties affecting the restframe $I$-band
Hubble diagram are discussed. Some sources are identified, for
instance the different methods applied for fitting the low and the
high redshift samples, selection effects for bright objects due to the
limiting magnitude of the search campaign, as well as
uncertainties in the $k$-correction calculations due to the presence
of the Ca IR triplet feature in the near infrared region of the SN
spectra. However, these systematic uncertainties differ from the ones
that could affect the restframe $B$-band Hubble diagram.

Restframe $I$-band observations of distant SNe~Ia are feasible, useful
and complementary to the already well established observations in the
$B$-band.

\begin{acknowledgements}
S.N. is grateful to Brian Schmidt for useful discussions on
$k$-corrections. We acknowledge the anonymous referee for useful
comments. Part of this work was supported by a graduate student grant
from the Swedish Research Council. AG is a Royal Swedish Academy
Research Fellow supported by a grant from the Knut and Alice
Wallenberg Foundation. This work was supported in part by the
Director, Office of Science, Office of High Energy and Nuclear
Physics, of the U.S. Department of Energy under Contract
No. DE-AC03-76SF00098.  Support for this work was provided by NASA
through grant HST-GO-08346.01-A from the Space Telescope Science
Institute, which is operated by the Association of Universities for
Research in Astronomy, Inc., under NASA contract NAS 5-26555.
\end{acknowledgements}

%
%
%
%

\end{document}